\documentclass[11pt,sort&compress]{elsarticle}
\makeatletter
  \long\def\pprintMaketitle{\clearpage
  \iflongmktitle\if@twocolumn\let\columnwidth=\textwidth\fi\fi
  \resetTitleCounters
  \def\baselinestretch{1}%
  \printFirstPageNotes
  \begin{center}%
 \thispagestyle{pprintTitle}%
   \def\baselinestretch{1}%
    {\large\bf\@title}\par\vskip5pt
    \normalsize\elsauthors\par\vskip5pt
    \footnotesize\itshape\elsaddress\par\vskip10pt
    \end{center}%
  \gdef\thefootnote{\arabic{footnote}}%
  }
\makeatother

\newcommand\blfootnote[1]{%
  \begingroup
  \renewcommand\thefootnote{}\footnote{#1}%
  \addtocounter{footnote}{-1}%
  \endgroup
}

\journal{}
\usepackage{amssymb}
\usepackage{amsmath}
\usepackage{amsfonts}
\usepackage{graphicx}
\usepackage{xspace}
\usepackage{epsf}
\usepackage{setspace}
\usepackage{listings}
\usepackage{float}
\usepackage{abstract}
\usepackage{outlines}
\usepackage{tikz}
\usepackage{algpseudocode}
\usepackage{algorithm}
\usepackage{longtable}
\usepackage{booktabs}
\usepackage{physics}
\usepackage{upgreek}

\usepackage[]{siunitx}
\DeclareSIUnit[quantity-product = ]\percent{\%} 

\usepackage[]{xcolor}
\usepackage{bm}
\newcommand{\ve}[1]{\bm{#1}}

\newcommand{\bu}{\ve{u}}

\newcommand{\bT}{\ve{T}}

\newcommand\Rey{\mbox{\text{Re}}\xspace}
\newcommand\Ma{\mbox{\text{Ma}}\xspace}
\newcommand\St{\mbox{\text{St}}\xspace}
\newcommand\Wo{\mbox{\text{Wo}}\xspace}
\newcommand\Kc{\ensuremath{\mathrm{K_c}}\xspace}

\newcommand\transpose{\mathsf{T}}

\newcommand{\rme}{\mathrm{e}}
\newcommand{\rmi}{\mathrm{i}}

\usepackage{xparse}
\DeclareDocumentCommand{\diff}{O{} m}{
	\frac{\mathrm{d} #1}{\mathrm{d}#2}
}
\DeclareDocumentCommand{\difftwo}{O{} m}{
	\frac{\mathrm{d}^2 #1}{\mathrm{d}#2^2}
}
\DeclareDocumentCommand{\pdiff}{O{} m}{
	\frac{\partial #1}{\partial #2}
}
\DeclareDocumentCommand{\pdifftwo}{O{} m}{
	\frac{\partial^{2} #1}{\partial #2^{2}}
}
\DeclareDocumentCommand{\integral}{O{} O{} m O{x}}{
	\int_{#1}^{#2} #3\ \mathrm{d}#4
}

\definecolor{lightblue}{rgb}{0.63, 0.74, 0.78}
\definecolor{seagreen}{rgb}{0.18, 0.42, 0.41}
\definecolor{orange}{rgb}{0.85, 0.55, 0.13}
\definecolor{silver}{rgb}{0.69, 0.67, 0.66}
\definecolor{rust}{rgb}{0.72, 0.26, 0.06}
\definecolor{purp}{RGB}{68, 14, 156}

\colorlet{lightrust}{rust!50!white}
\colorlet{lightorange}{orange!25!white}
\colorlet{lightlightblue}{lightblue}
\colorlet{lightsilver}{silver!30!white}
\colorlet{darkorange}{orange!75!black}
\colorlet{darksilver}{silver!65!black}
\colorlet{darklightblue}{lightblue!65!black}
\colorlet{darkrust}{rust!85!black}
\colorlet{darkseagreen}{seagreen!85!black}

\definecolor{c1}{rgb}{0.267,0.004,0.329}
\definecolor{c2}{rgb}{0.283,0.141,0.458}
\definecolor{c3}{rgb}{0.254,0.265,0.530}
\definecolor{c4}{rgb}{0.207,0.372,0.553}
\definecolor{c5}{rgb}{0.164,0.471,0.558}
\definecolor{c6}{rgb}{0.129,0.566,0.551}
\definecolor{c7}{rgb}{0.134,0.659,0.517}

\usepackage{lipsum}
\setlipsum{%
  par-before = \begingroup\color{gray},
  par-after = \endgroup
}

\usepackage[]{microtype}
\usepackage[margin=1in]{geometry}

\usepackage{stmaryrd}
\SetSymbolFont{stmry}{bold}{U}{stmry}{m}{n}

\usepackage{hyperref}
\hypersetup{
  colorlinks=true,
}

\usepackage[labelfont=bf,font=footnotesize]{caption}

\usepackage[nameinlink]{cleveref}
\crefname{lstlisting}{listing}{listings}
\Crefname{lstlisting}{Listing}{Listings}
\crefname{equation}{}{}

\usepackage[parfill]{parskip}

\begin{document}

\hypersetup{
  linkcolor=darkrust,
  citecolor=seagreen,
  urlcolor=darkrust,
  pdfauthor=author,
}

\begin{frontmatter}

 \title{{\large\bfseries Energy dissipation mechanisms in an acoustically-driven slit}}

\author[gtae]{Haocheng~Yu\corref{cor1}}
\ead{haochey@gatech.edu}
\author[gtcse]{Tianyi~Chu\corref{cor1}}
\ead{tchu72@gatech.edu}
\author[gtae,gtcse,gtme]{Spencer~H.~Bryngelson}

\cortext[cor1]{Equal contribution.}

\address[gtae]{Daniel Guggenheim School of Aerospace Engineering, Georgia Institute of Technology, Atlanta, GA 30332, USA\vspace{-0.15cm}}
\address[gtcse]{School of Computational Science \& Engineering, Georgia Institute of Technology, Atlanta, GA 30332, USA\vspace{-0.15cm}}
\address[gtme]{George~W.~Woodruff School of Mechanical Engineering, Georgia Institute of Technology, Atlanta, GA 30332, USA}

\date{}

\end{frontmatter}

\blfootnote{
\noindent Code available at: \url{https://github.com/MFlowCode/MFC}
}

\begin{abstract}
We quantify how incident acoustic energy is converted into vortical motion and viscous dissipation for a two-dimensional plane-wave passing through a slit geometry. 
We perform direct numerical simulations over a broad parameter space in incident sound pressure level (ISPL), Strouhal number ($\St$), and Reynolds number ($\Rey$). 
Spectral proper orthogonal decomposition (SPOD) yields energy-ranked coherent structures at each frequency, from which we construct mode-by-mode fields for spectral kinetic energy (KE) and viscous loss (VL) components to examine the mechanisms of acoustic absorption.
At $\mathrm{ISPL} = \SI{150}{\decibel}$, the acoustic--hydrodynamic energy conversion is highest when the acoustic displacement amplitude is comparable to the slit thickness, corresponding to a Keulegan--Carpenter number of order unity. 
In this regime, the oscillatory boundary layer undergoes periodic separation, resulting in vortex shedding that dominates acoustic damping.
VL accounts for \SIrange[]{20}{60}{\percent} of the KE contribution.
For higher acoustic frequencies, the confinement of the Stokes layer produces X-shaped near-slit modes, reducing the total energy input by approximately \SI{50}{\percent}. 
The influence of $\Rey$ depends on amplitude.
At $\mathrm{ISPL} = \SI{150}{\decibel}$, larger $\Rey$ values correspond to suppressed broadband fluctuations and sharpened harmonic peaks.
At $\mathrm{ISPL} = \SI{120}{\decibel}$, the boundary layers remain attached, vortex shedding is weak, absorption monotonically scales with viscosity, and the $\Rey$- and $\St$-dependencies become comparable.
Across all conditions, more than \SI{99}{\percent} of the VL is confined to a compact region surrounding the slit mouth.
The KE--VL spectra describe parameter regimes that enhance or suppress acoustic damping in slit geometries, providing a physically interpretable basis for acoustic-based design.

\end{abstract}

\section{Introduction}
\label{s:introduction}

The interaction between acoustic waves and fluid flows in confined geometries is a fundamental topic in fluid mechanics, with significant implications for acoustic damping, noise control, and the design of acoustic liners. 
Slit resonators, characterized by narrow openings in rigid boundaries, serve as canonical models for studying such interactions due to their simplicity and the richness of the underlying physics.
The vortices shed in these systems can exhibit irregular dynamics, despite the associated velocity fluctuations remaining small and the acoustically induced flow typically operating at low Mach and Reynolds numbers.
These systems exhibit complex behaviors from the coupling between incident acoustic waves and the generation of vortical structures, leading to energy conversion and dissipation mechanisms that are not fully understood. 
The canonical configuration, an infinitely long slit cut into a rigid plate and terminated by a deep backing cavity, was first analyzed by \citet{ingard1953theory} and later revisited~\citep{tam2000microfluid,tam2001numerical, tam2005computational, tam2008numerical}.
However, the damping of acoustic waves from their propagation through the slit, separate from the resonant response of the backing cavity, is usually considered only implicitly and has rarely been investigated as an isolated mechanism.

\begin{figure}[!ht]
    \centering
    \includegraphics[]{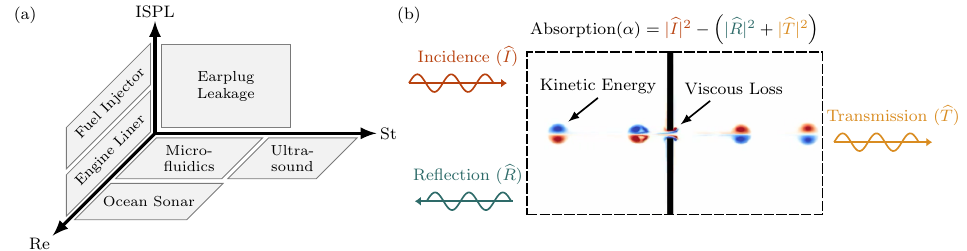}
    \caption{Schematics of (a) the parameter space explored in this study and (b) the global energy pathways, showing the conversion of incident acoustic energy into the kinetic energy of shed vortices and its subsequent viscous dissipation.}
    \label{fig:motivation} 
\end{figure}

Three nondimensional parameters characterize the acoustic source and the surrounding medium and play central roles in the engineering design of narrow slits, leaks, and small apertures: the acoustic amplitude, characterized by the incident sound pressure level (ISPL); the acoustic frequency, characterized by the Strouhal number ($\St$); and the viscous effects, characterized by the Reynolds number ($\Rey$). 
These effects are important in applications such as acoustic liners in jet engines, ocean sonars, hearing protection devices, and small structural gaps in transportation and aerospace systems; see the schematic in \cref{fig:motivation}~(a).
Depending on the application and the surrounding medium, one may seek either to attenuate the incident acoustic waves or to preserve them.

In the low-amplitude, low-frequency regime, the dissipative behavior of a slit can be predicted reasonably well by linearized boundary-layer theory, which attributes losses to viscous diffusion within the Stokes layer adjacent to the walls.
\citet{abily2023non} recently demonstrated that a transfer-matrix-method-based analytical model accurately predicts the sound absorption observed experimentally for amplitudes below $\SI{120}{\decibel}$ and frequencies below $\SI{2}{\kilo\hertz}$; 
In parallel, \citet{qu2023broadband} and \citet{hoppen2023helmholtz} proposed analytical models for Helmholtz resonators that remain in good agreement with both simulations and experiments up to \SI{140}{\decibel}.
For higher forcing levels, the flow inside the slit develops into a vortex roll-up, skew-symmetric jetting, and eventually, vortex shedding from the slit exit.
These effects emerge when the acoustic frequency and amplitude yield a Stokes boundary layer thickness comparable to the slit height, or when the acoustic particle displacement exceeds the slit thickness, corresponding to a Keulegan--Carpenter number $\Kc$ of order unity or greater.
In this regime, the kinetic energy of the induced vortices, together with their viscous dissipation, becomes the primary mechanism of acoustic energy loss~\citep{howe1980dissipation, cummings1984acoustic, tam2001numerical}.
Such nonlinear phenomena can substantially increase or decrease the acoustic resistance of slit-based resonant elements relative to their linear prediction~\citep{tam2001numerical,tam2005computational}, and they remain a major source of uncertainty in practical design guidelines.

To characterize the dependence on Reynolds number, Strouhal number, and ISPL, we construct a database for a two-dimensional slit configuration with anechoic termination using direct numerical simulations (DNS).
This database is suitable for examining the onset and scaling of sound-induced vortex shedding, the spatial structure of kinetic-energy-dominant and viscous-loss-dominant components, and the role of nonlinear acoustic response in narrow-slit geometries.
We utilize this database to clarify the underlying dissipation mechanisms of plane acoustic waves as they propagate through the slit.


While the existence of vortex-based energy conversion in slit resonators has been well-examined using probes, fundamental scientific questions of engineering relevance remain open.
The $\St$-dependence remains to be characterized.
Although empirical evidence indicates that the interaction between the oscillating boundary layer and the bulk flow is strongest when $\St$ is $O(1)$~\citep{tam2005computational,tam2008numerical,chen2020microscopic,aulitto2022effect}, the scaling of modal energy across different spectral components with $\St$ has not been quantified.
Amplitude dependence also requires further clarification.
Most available measurements collapse acoustic amplitude into a single nondimensional excursion parameter, such as absorption coefficient and acoustic impedance~\citep{chung1980transfer,tam2001numerical,tam2005computational,abily2023non}; however, this practice can mask the cascade of triadic interactions that emerges once the particle displacement exceeds roughly \SI{10}{\percent} of the slit height.
In addition, the $\Rey$-dependence is not fully resolved.
At fixed acoustic amplitude, increased Reynolds number in irregular regimes expands the inertia-dominated near-wall region and promotes earlier shear-layer roll-up near the slit inlet~\citep{abily2023non}.
The interaction of this shift with amplitude-induced vortex breakdown remains unclear, yet it influences the high-frequency performance limits of practical resonator designs.
To examine the energy dissipation mechanisms across these parameter variations, modal decomposition techniques provide a framework for quantifying the contributions of coherent flow structures across different scales.

Our goal is to disentangle the contributions of the acoustic component, quantified by scalar pressure fluctuations, from the kinetic energy (KE) component of the induced flow field, from vortex dynamics, together with the overall viscous dissipation, which we refer to as the viscous loss (VL) component.
Proper orthogonal decomposition(POD) provides an efficient statistical way to extract energy-ranked spatial structures from flow field data~\citep{lumley1967structure,lumley1970stochastic}.
In practice, POD uses the method of snapshots, which involves the eigendecomposition of the spatial cross-correlation tensor~\citep{sirovich1987turbulence}.
In the context of acoustic slits, POD has been applied to phase-locked particle image velocimetry (PIV) measurements of a dual-slit cavity, where the most energetic modes exhibit pronounced asymmetry~\citep{tang2025phase}.
In general, the POD modes do not distinguish between hydrodynamic and acoustic components, as the method lacks scale separation in frequency~\citep{berkooz1993proper}.
Dynamic mode decomposition (DMD) addresses part of this limitation by providing a spatiotemporal decomposition of time-resolved data streams, associating each mode with a single complex frequency and its corresponding growth/decay rate~\citep{schmid2010dynamic}.
Using DMD on DNS data of transient vortex dynamics in an acoustic-driven slit, \citet{qiang2022aeroacoustic} showed that the most dominant dynamical structures also exhibit asymmetry perpendicular to the direction of sound propagation.
Nonetheless, DMD can be sensitive to subsampling and may struggle to converge in statistically stationary turbulent flows, typically requiring carefully designed ensemble-based formulations.

As an alternative to the above data-driven modal decomposition techniques, spectral POD (SPOD) operates analogously to POD but in the frequency domain, enabling the identification of energy-ranked structures that evolve coherently in both space and time~\citep{schmidt2018spectral,towne2018spectral}.
SPOD modes represent both the statistical and dynamical content of the flow at each frequency, optimally representing second-order statistics and serving as optimally averaged ensemble DMD modes~\citep{towne2018spectral}.
Readers are referred to \citet{schmidt2020guide} for the practical implementation of SPOD.
SPOD has been widely applied to extract both coherent acoustic and hydrodynamic structures in various turbulent flows, including cylinder~\citep{awasthi2025coherent}, airfoil wakes~\citep{sano2019trailing, himeno2021spod}, and jets~\citep{schmidt2018spectral,nekkanti2021modal,nogueira2022wave,bugeat2024acoustic}.
In these contexts, SPOD isolates dominant spectral flow features, such as vortex-shedding modes in resonators. 
It provides a natural way to analyze their associated linear mechanisms, including advection, production, and dissipation.
Recently, \citet{tang2023combined} proposed an SPOD-based noise reduction strategy to improve the performance of acoustic liners inside a steam turbine control valve through three-dimensional delayed detached eddy simulation (DDES).
In this work, we use SPOD to separate and quantify the mode-by-mode, frequency-resolved KE and VL components and their spatial structures, providing explicit modal-spectral energetics in a slit mouth configuration.
\Cref{fig:motivation} (b) shows a schematic of how the total damping of the incident acoustic energy within a control volume is distributed among different spectral KE and VL components.
The resulting spectral KE and VL fields exhibit localized structures within and near the slit, through which we reveal the acoustic--KE--VL energy-exchange mechanism at the slit mouth as a coupled effect of $\St$, $\Rey$, and ISPL.
By spatially and spectrally integrating these fields, we recover conventional global measures of acoustic power loss inferred from reflection and transmission, thereby linking the underlying modal energetics to the net absorption.

This paper is outlined as follows.
In \cref{sec:DNS}, we introduce the high-fidelity numerical approach for a plane wave traversing a two-dimensional slit mouth, and we detail the resulting database across the parameter space in \cref{sec:Database}.
In \cref{sec:spectral}, we use SPOD-based spectral analysis to quantify the dissipation mechanisms associated with each spectral KE and VL component and demonstrate how variations in $\St$, $\Rey$, and ISPL affect the energy transfer from acoustic waves to vortical structures.
Our results reveal the acoustic--KE--VL energy-exchange mechanism at the slit mouth, its dependence on the governing parameters, and its connection to the onset of boundary-layer separation.
We demonstrate budget closure by comparing volume-integrated dissipation with the independently measured acoustic absorption coefficient.
\Cref{sec:Discussion,sec:Conclusion} summarizes the main contributions and discusses limitations and implications for acoustic slit design.

\section{Direct Numerical Simulation}
\label{sec:DNS}

Throughout this work, we adopt the following notational convention:
scalar-valued functions are denoted by italicized symbols (e.g., $q$);
vector-valued functions by bold italic symbols (e.g., $\boldsymbol{q}$).
Spatially discretized quantities are denoted with vector notation, discretized scalars $\vec{{q}}$, and discretized vectors as $\vec{\mathbf{q}}$.

\subsection{Governing equations and numerical method}\label{subsec:eqns}
To represent the propagation of sound waves through two-dimensional slits, the flow motion is modeled using the compressible Navier--Stokes equations, encompassing the continuity, momentum, and energy equations,
\begin{align}
    \pdv{\rho}{t} + \grad\cdot \left(\rho \vb*u  \right)&=h_s h_{\delta}/c,\label{cont_eqn}\\
    \pdv{}{t} \left(\rho \vb*u \right) + \grad \vdot \left( \rho\vb*{u} \vb*u^\transpose \right) + 
        \grad p- \grad\vdot\vb*{T} &=h_s h_{\delta} \vb*{n},\label{mom_eqn}\\
    \pdv{}{t} \left(\rho E\right) + \grad \cdot \left[ (\rho E + p)\bu -  \bT \cdot \bu \right] &= 0,
\end{align}
where $\vb*u = [u,\;v]^\transpose$ is the velocity vector field, $c$ is the speed of sound, $\vb*{n}$ is the unit vector directed from the sound source toward the slit, and
\begin{gather}
    \vb*{T} \equiv\frac{1}{\Rey} \left[\grad \vb*{u}+(\grad \vb*{u})^\transpose -\frac{2}{3}\left(\grad\vdot \vb*{u}\right)\vb*{I}\right]
    \label{eq:viscous_stress}
\end{gather} 
is the viscous stress tensor, with the bulk viscosity assumed negligible and thus omitted.
The one-way plane wave source is modeled using two nonconservative source terms~\citep{maeda2017source}, where $h_s$ is the time-dependent amplitude of the monopole and dipole sources, and $h_{\delta}$ represents the Gaussian monopole support function provided by~\citet{maeda2017source}.
The source terms in equations \cref{cont_eqn} and \cref{mom_eqn} represent a planar monopole and dipole source, respectively.
The source terms cancel out each other in one direction and thus result in a one-way planar sound source.
All sound waves are assumed to be harmonic, implying that the time-dependent amplitude,
\begin{align}
    h_s = \widehat{A}\sin(2\pi f_s t),
\end{align}
is sinusoidal with time, where $\widehat{A}$ is the pressure amplitude and $2\pi f_s$ is the angular frequency of the sound source.
Closure is achieved using the ideal gas equation of state,
\begin{align}
    p = \left(\gamma - 1\right) \rho e, \label{eq:gas}
\end{align}
which relates the internal energy, $e$, to the total energy of the fluid, $E = e + \|\vb*{u}\|^2/2$.
Here, $\gamma=1.4$ is the ratio of specific heats.
Equations \crefrange{cont_eqn}{eq:gas} are nondimensionalized by the speed of sound, $c = \SI{343}{\meter \per \second}$ and the slit opening and thickness $d = \SI{0.8}{\milli \meter}$, and are parameterized by the Strouhal number $\St = f_{s} d / c$ and
the Reynolds number $\Rey=\rho c d/\mu$.
Here, $\rho$ is the density of air and $\mu$ is the dynamic viscosity, which varies with the Reynolds number across different simulations.
The nominal Strouhal number and Reynolds number are defined as $\St_0 = f_{s,0} d / c$ and $\Rey_0 = \rho_0 c d / \mu_0$, respectively, where $f_{s,0} = \SI{0.5}{\kilo\hertz}$ is the reference frequency.
The reference density and dynamic viscosity are taken as near-sea-level air properties, $\rho_0=\SI{1.19}{\kilogram\per \cubic\meter}$ and $\mu_0 = \SI{1.84e-5}{\pascal \second}$, respectively.
All the length scales, $x$, are nondimensionalized by $d$, time $t$ by $c/d$, velocity $u$ by $c$, pressure $p$ by $\rho_0 c^2$, the source terms in the \cref{cont_eqn} $h_s h_{\delta}/c$ by $d/\rho_0c$, and source terms in the \cref{mom_eqn} $h_s h_{\delta} \vb*n$ by $d/\rho_0c^2$.

We perform two-dimensional (2D) direct numerical simulation simulations using MFC, a GPU-accelerated compressible flow solver~\citep{bryngelson2021mfc, radhakrishnan2024method, wilfong2025mfc, wilfong-hpctests, wilfongGB25}.
Equations \crefrange{cont_eqn}{eq:viscous_stress} can be written compactly as
\begin{gather}
    \frac{\partial \vb*{q}}{\partial t} + \grad \cdot \vb*{F}(\vb*{q}) = \vb*{s}(\vb*{q}),
    \label{eq1}
\end{gather}
where,
\begin{align}
    \vb*{q} &= \begin{bmatrix}
        \rho \\
        \rho \vb*{u} \\
        \rho E \\
    \end{bmatrix}, \quad
    \vb*{F} = \begin{bmatrix}
        \rho \vb*{u} \\
        \rho \vb*{u}\otimes \vb*{u}   + p\vb*{I} - \vb*{T} \\
        (\rho E + p) \vb*{u} - \vb*{T} \cdot \vb*{u} \\
    \end{bmatrix}, \quad \text{and} \quad  
    \vb*{s} = \begin{bmatrix}
        h_s h_{\delta}/c \\
        h_s h_{\delta} \vb*n \\
        0 \\
    \end{bmatrix},
    \label{eq2}
\end{align}
represent the state vector, flux tensor, and the source vector, respectively.
The spatial discretization of equations \cref{eq1,eq2} uses a finite volume method, which is well-suited for problems involving irregular geometries and complex compressible flows. 
A fifth-order accurate Weighted Essentially Non-Oscillatory~(WENO) scheme~\citep{coralic2014finite} is used to reconstruct the fluxes and viscous terms, which handles sharp gradients in the high-amplitude solution without introducing spurious oscillations.
We perform temporal discretization using a total variation diminishing third-order accurate Runge--Kutta scheme~\citep{Gottlieb1998}.

\subsection{Simulation Configurations} \label{subsec:sim_configuration}

\begin{figure}[!ht]
    \centering
    \includegraphics[]{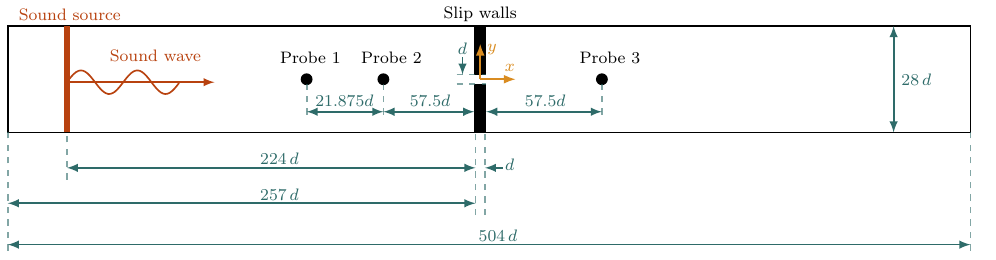}
    \caption{Schematic of the flow configuration in the $x$--$y$ domain simulated in this work.}
    \label{fig:dns_config} 
\end{figure}

\Cref{fig:dns_config} shows the configuration of the computational domain, $\Omega$, discretized with a uniformly structured mesh with equal grid spacing.
Rightward-going acoustic waves were emitted from a planar acoustic source $224 d$ from the incident face of the slit resonator ($28 d$ away from the left domain boundary), where $d = \SI{0.8}{\milli \meter}$.
Ghost-cell non-reflective boundary conditions are used on the left and right boundaries, where primitive variables in the ghost cells are extrapolated from the interior to satisfy $\partial(\cdot)/\partial n \approx 0$, and $n$ represents the direction perpendicular to the boundary.
In practical systems, the boundary layer on the top and bottom domain walls may contribute to the overall absorption.
However, the present simulations intentionally isolate dissipation associated with the slit mouth itself, so we neglected the dissipation of sound along the top and bottom domain boundaries and treated these surfaces as characteristic slip walls, following \citet{thompson1990}.
The no-slip condition was enforced on the incident face of the slit and all of its interior walls based on the immersed boundary method~\citep{tseng2003ghost}. 

We use the incident sound pressure level (ISPL) to represent the decibel sound level at a specific frequency, defined as
\begin{equation}
    \mathrm{ISPL} = 20 \log_{10}\left(\widehat{p}/\widehat{p}_\mathrm{ref}\right),
    \label{eq:ISPL}
\end{equation}
where $\widehat{p} = \widehat{A}/\sqrt{2}$ is the dimensional root-mean-squared pressure (RMS) of the sound, and $\widehat{p}_\mathrm{ref} = \SI{20}{\micro \pascal}$ is the standard reference pressure in air. 

The 2D slit under consideration is a wall-bounded problem, and the oscillatory boundary layer is a near-wall feature characterized by steep gradients, rather than a propagating wave.
Previous work has found that the acoustic-driven flow bounded by small slit resonators, as small as $\SI{1}{\cm}$, is laminar at a high incident level of $\SI{155}{\decibel}$~\citep{tam2000microfluid}.
The laminar condition with an oscillatory wave gives rise to a viscous Stokes layer inside the slit opening whose thickness is much smaller than the acoustic wavelength of our range of interest~\citep{tam2005computational, tam2010computational}.
The wavelength of the oscillatory Stokes layer $\lambda_s$ is defined as
\begin{align}
        \lambda_s \equiv 2 \pi \delta_s= 2 \sqrt{{\pi\nu}/{f_s}}
    \label{eq:Stokes_layer}
\end{align}
where $\delta_s$ is the thickness of the Stokes layer, $\nu = \mu/\rho$ is the kinematic viscosity, and $f_s$ is the acoustic frequency~\citep{white1991viscous,tam2005computational}.
For varying acoustic frequencies, simulations are performed using at least ten cells per Stokes layer wavelength ($\lambda_s / \Delta_x \geq 10$) to ensure negligible numerical errors when the Stokes boundary layers are resolved, where $\Delta_x$ is the grid spacing.
Grid convergence is confirmed by the negligible change in simulated acoustic power absorption coefficients when the cell count is doubled.
Simulations are performed over multiple acoustic cycles with a constant time step to ensure the system reaches a dynamic steady state before data collection.

\subsection{DNS Verification} \label{subsec:sim_validation}

\begin{figure}[htb]
    \centering
    \includegraphics[]{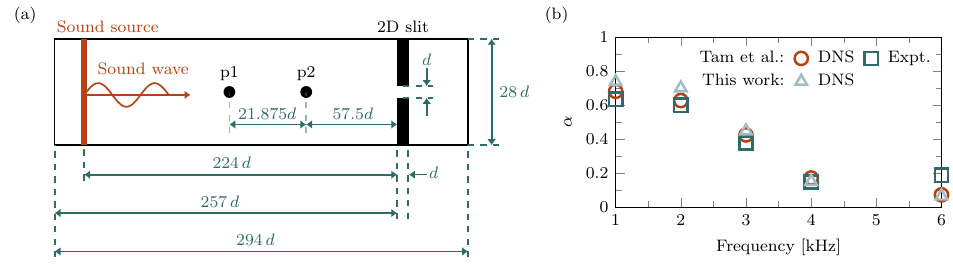}
    \caption{A 2D slit resonator for verification showing (a) the computational domain with acoustic source (not to scale) and (b) comparison of power absorption coefficient spectra among direct numerical simulation (DNS) and experiment (Expt.) of \citet{tam2001numerical} with discrete tones at $\mathrm{ISPL}= \SI{150}{\decibel}$.}
    \label{dns_validation} 
\end{figure}

We verify our numerical simulations using a 2D simulation of a discrete tone ($\mathrm{ISPL} = \SI{150}{\decibel}$) propagating as a plane wave through a 2D slit resonator.
The opening and thickness of this slit, $d$, are scaled to be $1/28$ of the domain height to reproduce the configuration of \citet{tam2001numerical}.
A similar verification simulation has been conducted by \citet{yu2024numerical, yu2025transmission}.
This verification setup is analogous to the cases examined in this study, with the computational domain having a nondimensional length of $294d$, as illustrated in \cref{dns_validation}~(a).
In this verification case, we use non-reflective boundary conditions at the left horizontal boundary and impermeable slip walls for the remaining domain boundaries.
The simulation time and grid size match those of the 2D cases in \cref{table:dns_config}.
Thus, the only differences between this verification simulation and the 2D cases in \cref{table:dns_config} are the termination boundary condition and the $x$-direction extent of the computational domain.

We use the power absorption coefficient as the metric to quantify the acoustic performance.
The absorption coefficient that represents the percentage of the sound power dissipated by the slit resonator is defined as
\begin{equation}
    \alpha \equiv 1 - |\widehat{R}|^2 - |\widehat{T}|^2,
    \label{eq:acoustic_alpha}
\end{equation}
where $|\widehat{R}|^2$ and $|\widehat{T}|^2$ are the power reflection and transmission coefficients, respectively.
Following \citet{leung2007duct} and \citet{tam2005computational}, these coefficients are calculated from the recorded time histories using the transfer function method of \citet{chung1980transfer}.
In the present verification case with a reflective termination, all the transmitted waves will be reflected by the termination, yielding $|\widehat{T}|^2 \approx 0$.
The power absorption coefficient is approximated as $\alpha \approx 1 - |\widehat{R}|^2$.
For discrete-tone analysis, the power reflection coefficient at a given frequency is determined via linear interpolation between the two adjacent spectral points that bracket the target frequency.

We collect pressure data at two points along the horizontal center of the domain at each simulation time step. 
The two points are $21.875d$ and $57.5d$ upstream from the slit incident face, where higher-order modes are evanescent.
The sampling frequency of the recorded time history is as large as the inverse of the simulation time step.
\Cref{dns_validation}~(b) shows a comparison of the present absorption coefficient spectra with those reported by \citet{tam2001numerical} at a nominal incident sound intensity of \SI{150}{\decibel}.
The good agreement between our results and the baseline demonstrates the accuracy of the present numerical method in simulating slit resonators under acoustic excitation.
In the following, the MFC solver is used to generate a comprehensive database for the acoustic slit, encompassing a wide range of parameters as detailed in \cref{sec:Database}.

\section{Database of 2D acoustic slit}
\label{sec:Database}

\subsection{Setup}
\label{subsec:Database_setup}

The dominant factors influencing the performance of the acoustic slit include incident sound pressure level, acoustic frequency, and Reynolds number.
To isolate the effect of each parameter, simulations are performed across various combinations of these three variables.
Specifically, two ISPLs, $\SI{120}{\decibel}$ and $\SI{150}{\decibel}$, are considered to represent cases of weak and strong sound excitations, respectively.
The acoustic frequency is varied from $\SI{0.5}{\kilo\hertz}$ to $\SI{6}{\kilo\hertz}$ to represent a broad range of acoustic behaviors.
For a fixed speed of sound and slit width, three Reynolds numbers are examined to investigate the effect of the varying dynamic viscosity.
\Cref{table:dns_config} summarizes the considered parameters and the corresponding simulation configurations.

\begin{table}[htb]
    \centering
    \caption{
        Simulation configuration of seven different acoustic frequencies.
        The following simulations include cases with three Reynolds numbers, $\Rey/\Rey_0 = 1$, $1/2$, and $1/3$, and two sound pressure levels: $\mathrm{ISPL} = \SI{120}{\decibel}$ and $\SI{150}{\decibel}$.
        The three varying parameters result in 42 unique combinations.
        The total number of cells in the $x$-direction ($N_x$) and the $y$-direction ($N_y$) describes the grid size of the computational domain.
        The number of cells per Stokes' wavelength $\lambda_s$ describes the cell size. 
        We use the number of time steps per acoustic cycle to describe the time step size. 
        \label{table:dns_config}
        }
    {\small
    \begin{tabular}{S[table-format=1.1] r r r r r r}
        \toprule
        \multicolumn{1}{c}{Frequency} &
        \multicolumn{1}{c}{$\St / \St_{0}$} &
        \multicolumn{1}{c}{Domain Size} &
        \multicolumn{1}{c}{Cell Size} &
        \multicolumn{1}{c}{Time Step} &
        \multicolumn{1}{c}{Duration} &
        \multicolumn{1}{c}{Analysis Range} \\
        \multicolumn{1}{c}{(\si{\kilo\hertz})} &
        \multicolumn{1}{c}{(---)} &
        \multicolumn{1}{c}{($N_x\times N_y$)} &
        \multicolumn{1}{c}{($\lambda_s/\Delta_x$)} &
        \multicolumn{1}{c}{($10^4$/cycle)} &
        \multicolumn{1}{c}{(cycles)} &
        \multicolumn{1}{c}{(cycles)} \\
        \midrule
       0.5 & 1  &$20160 \times 1120$ & 30 & $8.0$  & 25  & 10--25\\ 
        1  & 2  &$20160 \times 1120$ & 20 & $4.0$  & 50  & 19--50\\ 
        2  & 4  &$20160 \times 1120$ & 15 & $2.0$  & 75  & 12--75\\ 
        3  & 6  &$20160 \times 1120$ & 12 & $1.3$  & 113 & 18--113\\ 
        4  & 8  &$20160 \times 1120$ & 10 & $1.0$  & 150 & 23--150\\ 
        5  & 10 &$25200 \times 1400$ & 12 & $0.8$  & 139 & 20--139\\ 
        6  & 12 &$25200 \times 1400$ & 10 & $0.7$  & 167 & 24--167\\
        \bottomrule
    \end{tabular}
    }
\end{table}

All simulations were performed using MFC on the NCSA~Delta supercomputer.
Each case was executed on three NVIDIA~A100 GPUs.
The overall GPU utilization exceeds \SI{90}{\percent}~\citep{radhakrishnan2024method}.
The required wall time of each simulation is approximately \SI{40}{\hour} and depends weakly on the acoustic frequency and the corresponding grid resolution.
Numerical stability is verified under a Courant--Friedrichs--Lewy number (CFL) of less than $0.5$.
For the simulations documented in \cref{table:dns_config}, the raw data of the analyzed cycles are made publicly available through the Georgia Tech Digital Repository, comprising 42~MATLAB-formatted \texttt{mex} files.
For each case, the full velocity and density fields are stored over the analysis range indicated in the table, comprising 640 time snapshots.
The public flow field data are down-sampled by a factor of 2 in each coordinate direction.
Each file is approximately \SI{4.5}{\giga\byte} in size and corresponds to a unique parameter combination.

For the configurations listed in \cref{table:dns_config}, the acoustically induced flow remains at low Mach number ($\Ma$), with $\Ma \leq 0.05$ throughout the computational domain.
Velocities and their fluctuations are thus small compared with the ambient speed of sound.
Consequently,  compressibility effects in the induced flow are weak, and the pressure fluctuations associated with acoustic propagation are only mildly affected by local density variations, which remain below \SI{0.5}{\percent} of their initial value. This behavior is consistent with previous studies of low-Mach-number oscillatory flows in resonant cavities, which have shown that the velocity field is predominantly solenoidal and that compressibility enters only as a higher-order correction~\citep{landau1987fluid}.
Despite the relatively low Mach numbers, the flow outside the slit exhibits complex, irregular dynamics, whereas the flow inside the slit remains laminar.
We emphasise, however, that compressibility cannot be neglected in the simulations, as compressibility is essential for accurately resolving the propagation of acoustic waves.

\subsection{Data} \label{subsec:Database_data}

\begin{figure}[ht!]
    \centering
    \includegraphics[]{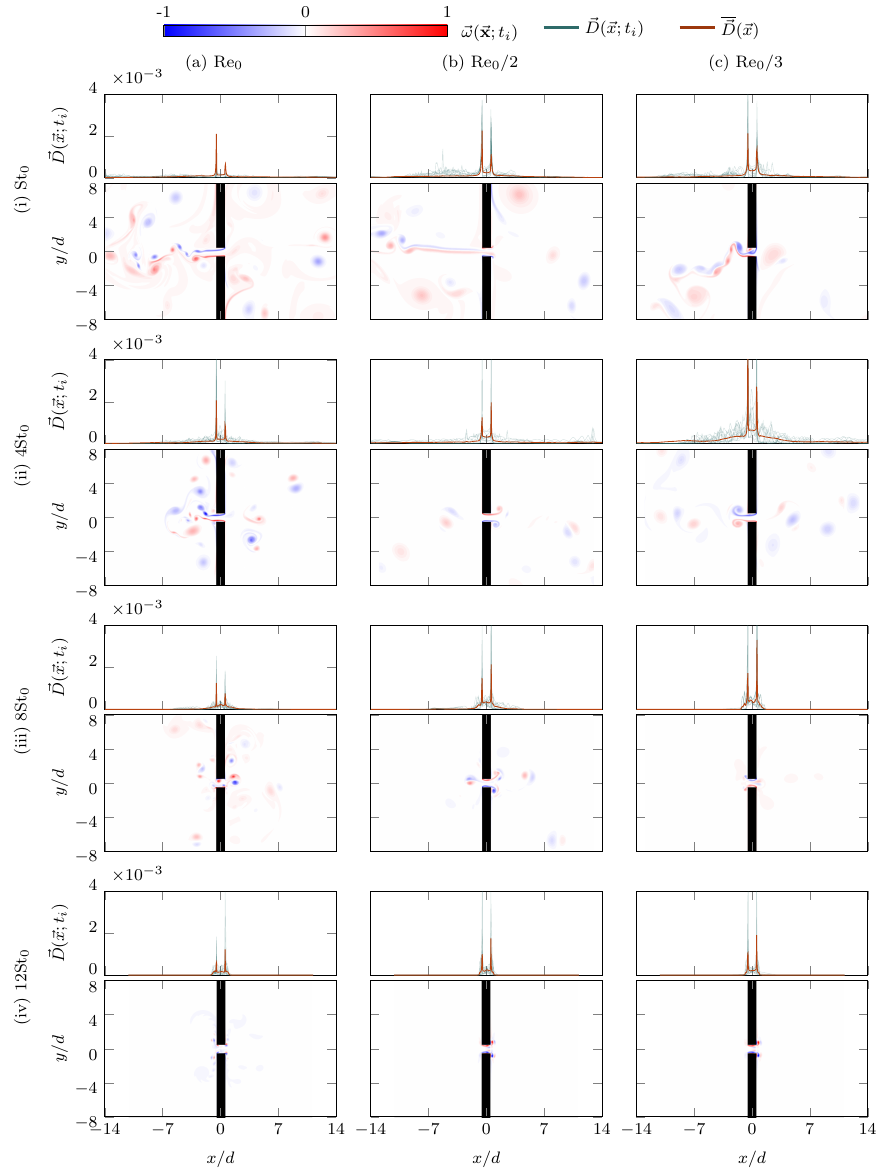}
    \caption{
     Instantaneous VL (integrated across the $y$ direction), $\vec{D}(\vec{x};t_i)$, and vorticity fields, $\vec{\omega}(\vec{\mathbf{x}};t_i)$, for different $\St$--$\Rey$
     combinations at $\mathrm{ISPL} = \SI{150}{\decibel}$.
     Temporal-averaged VL fields, \raisebox{-0.5ex}{$\smash{\overline{\vec{D}}(\vec{x})}$}, are shown for comparison.
     Cases in column (a)--(c) represent $\Rey=\Rey_0$, $\Rey_0/2$, and $\Rey_0/3$. Cases in row (i)--(iv) represent $\St=\St_0$, $4\St_0$, $8\St_0$,and $12\St_0$.
    }
    \label{instantaneous_vorticity_150db} 
\end{figure}

To illustrate the rich physics in the database, \cref{instantaneous_vorticity_150db} shows the normalized instantaneous vorticity fields,
\begin{align}\label{eqn:vorticity}
    \vec{\omega}(\vec{\mathbf{x}};t_i)
     = \frac{
        \left[\partial_x \vec{v}-\partial_y \vec{u}\right](\vec{\mathbf{x}};t_i)
    }{
        \max \{\left[\partial_x \vec{v}-\partial_y \vec{u}\right](\vec{\mathbf{x}};t_i) \}
    },
\end{align}
for various $\St$--$\Rey$ combinations at an incident sound pressure level of $\mathrm{ISPL} = \SI{150}{\decibel}$.
Here, $\vec{\mathbf{x}}$ denotes the discretized spatial coordinate vector over the computational domain.
Across all cases, the snapshots reveal well-resolved boundary layers generated as the flow interacts with the slit, followed by the formation of periodic vortex shedding.
These results indicate that a portion of the incident acoustic energy is converted into vortical motion through interactions with the two-dimensional slit.
These vortical structures intensify and periodically detach from the slit, forming coherent vortices that dominate the near-slit region.
Once shed, the vortices convect both upstream and downstream, undergoing complex interactions that give rise to a hierarchy of larger and smaller scales of motion.
As the acoustic frequency increases, the vortical structures gradually become confined to the vicinity of the slit, suppressing long-range interactions and limiting the generation of far-field vorticity.


To further quantify the viscous dissipation associated with these vortical structures, we consider the irreversible viscous dissipation rate $\vb*{T}:\nabla\vb*{u}$, which (up to compressible corrections) is non-negative and represents thermodynamic loss, and the viscous work density $\vb*{u}\cdot(\nabla\cdot\vb*{T})$.
In the current case, these two terms have a negligible difference, as the boundary flux,
\begin{gather}
    \int_{\partial\Omega}(\boldsymbol{u}\cdot\boldsymbol{T})\cdot\boldsymbol{n}\,\dd S,
\end{gather}
is negligible for the chosen analysis domain.
\Cref{instantaneous_vorticity_150db} further examines the instantaneous irreversible viscous loss (VL) fields integrated in the $y$-direction,
\begin{align}\label{eqn: VL_instant}
\vec{D}(\vec{x};t_i)=\sum_{y\in[-14d,14d]}
\left[ \vec{\mathbf{T}}: \grad \vec{\mathbf{u}}\right](\vec{\mathbf{x}};t_i)\Delta_y,
\quad
\text{with} \quad
\vec{\mathbf{T}}
=\frac{1}{\Rey}\Big(\grad\vec{\mathbf{u}}+\grad\vec{\mathbf{u}}^\transpose-\frac{2}{3}(\grad\cdot\vec{\mathbf{u}})\mathbf{I}\Big),
\end{align}
where $\Delta_y=\Delta_x$ under uniform grid spacing.
The corresponding temporal averages over $N_t$ snapshots are
\begin{align}
    \overline{\vec{D}}(\vec{x}) = \frac{1}{N_t} \sum_{i=1}^{N_t}{\vec{D}}(\vec{x};t_i).
\end{align}
Across all cases, the VL fields peak near the slit and attain values larger than those in other regions, indicating that the dominant energy loss is associated with the attached vortices.
In contrast, the detached vortices contribute comparatively little.
Interestingly, at the lower acoustic frequency of $\St = \St_0$, VL is larger near the inlet of the slit than at the outlet.
This trend is consistent with the corresponding flow fields, where more pronounced upward-traveling vortex shedding occurs at low frequencies.
At lower Reynolds numbers within the range considered, viscous effects become more pronounced relative to inertial ones, leading to greater VL while still sustaining vortex shedding.

\begin{figure}[ht!]
    \centering
    \includegraphics[]{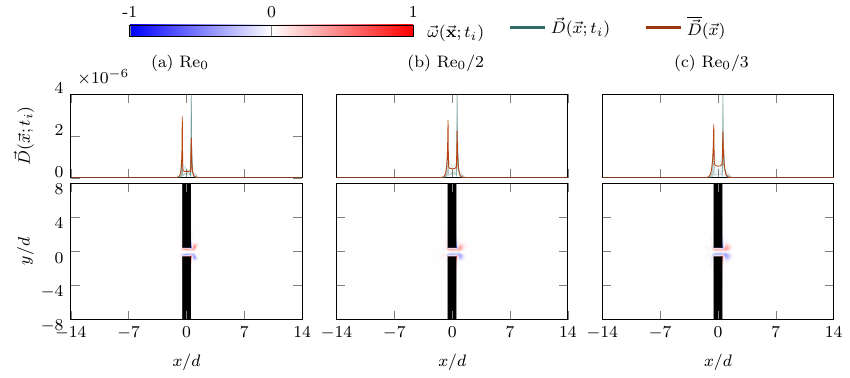}
    \caption{
    Instantaneous VL (integrated across the $y$ direction), $\vec{D}(\vec{x};t_i)$, and vorticity fields, $\vec{\omega}(\vec{\mathbf{x}};t_i)$, at $\mathrm{ISPL} = \SI{120}{\decibel}$ and $\St=\St_0$: (a) $\Rey=\Rey_0$; (b) $\Rey_0/2$; and (c) $\Rey_0/3$.
    }
    \label{instantaneous_vorticity_120db} 
\end{figure}

For comparison, \cref{instantaneous_vorticity_120db} shows the results at $\mathrm{ISPL} = \SI{120}{\decibel}$.
The flow fields with higher Strouhal numbers are similar, so we focus on a representative Strouhal number of $\St_0$.
At this weaker sound excitation, the vortex shedding remains nearly symmetric, with vorticity confined to boundary layers adjacent to the slit walls.
No evidence exists of large-scale vortex pairing or chaotic modulation.
Compared to high-intensity acoustic excitation, these cases exhibit substantially weaker VL by approximately three orders of magnitude.

\begin{figure}[htb]
    \centering
    \includegraphics[]{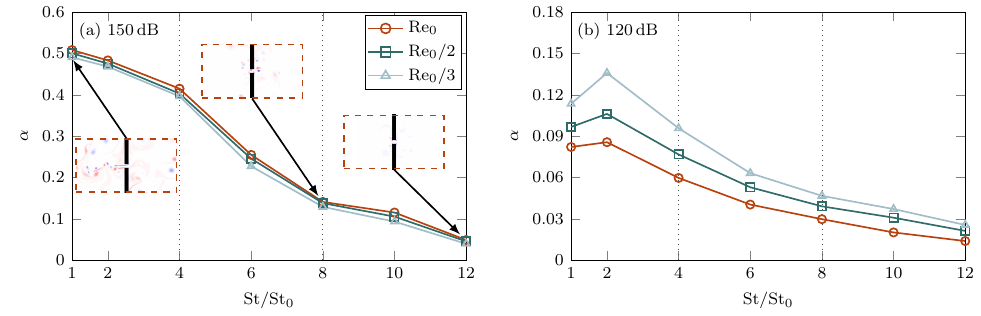}
    \caption{The power absorption coefficients $\alpha \equiv 1 - |\widehat{R}|^2 - |\widehat{T}|^2$ across all the $\St$--$\Rey$ combinations listed in \cref{table:dns_config}, subject to (a) $\mathrm{ISPL} = \SI{150}{\decibel}$ and (b) $\mathrm{ISPL} = \SI{120}{\decibel}$.
    Probe locations are given in \cref{fig:dns_config}.
    Insets in (a) show the normalized instantaneous vorticity fields, $\vec{\omega}(\vec{\mathbf{x}};t_i)$, of cases (a, i), (a, iii), and (a, iv) in \cref{instantaneous_vorticity_150db} to highlight the change in vorticity field across the range of $\alpha$ observed as $\St$ increases for $\mathrm{ISPL} = \SI{150}{\decibel}$ and $\Rey = \Rey_0$.
    }
    \label{dns_absorption} 
\end{figure}

Following the method outlined in \cref{subsec:sim_validation}, we compute and present the power absorption coefficient in \cref{dns_absorption} to compare the total energy absorption at varying parameter combinations.
At $\mathrm{ISPL} = \SI{150}{\decibel}$, for larger $\Rey$, the $\alpha$ is larger.
Power absorption at $\Rey = \Rey_0$ is approximately $0.01$ and $0.02$ larger than at $\Rey = \Rey_0/2$ and $\Rey = \Rey_0/3$ respectively.
On the contrary, for larger $\Rey$ (i.e. lower viscosity), $\alpha$ is smaller at $\mathrm{ISPL} = \SI{120}{\decibel}$.
Power absorption at $\Rey = \Rey_0$ is approximately $0.015$ and $0.03$ smaller than at $\Rey = \Rey_0/2$ and $\Rey = \Rey_0/3$ respectively.

The dependence of $\alpha$ on $\Rey$ remains significantly weaker, by at least an order of magnitude, than its dependence on $\St$ at $\mathrm{ISPL}=\SI{150}{\decibel}$, where the variation of $\alpha$ across the frequency sweep exceeds $0.4$. 
On the contrary, at $\mathrm{ISPL}=\SI{120}{\decibel}$, the $\Rey$-dependence becomes comparable to the $\St$-dependence, reflecting that in the absence of strong vortex shedding, the absorption is dominated by the viscous loss (VL), and not by frequency-dependent dynamics.
In sum, \cref{dns_absorption} shows two distinct absorption regimes: a vortex-dominated regime at high ISPL, where $\St$ is the primary dominating parameter, and a viscosity-dominated regime at low ISPL, where $\Rey$ plays a comparable role.

The analysis domain is confined to $\Omega = [x/d, y/d]\in [-14, 14] \times [-14, 14]$ for $\St/\St_0 \le 8$ and $\Omega = [x/d, y/d]\in [-11.2, 11.2] \times [-14, 14]$ for $\St/\St_0 \ge 10$. 
The horizontal distribution of VL identifies the region containing more than \SI{99}{\percent} of the total VL.
Within this region, more than \SI{99}{\percent} of the total kinetic energy (KE) associated with the induced flow field is retained, giving the spatial extent used henceforth for spectral analysis.

This database is directly relevant to real-world engineering design scenarios involving sound at different ISPLs and frequencies in different media with different viscosities.
The ISPL sets the magnitude of the imposed velocity oscillation through the plane wave approximation of acoustic impedance $u \approx p/\widehat{z}$, where $\widehat{z} = \rho c$ is the characteristic acoustic impedance of air.
The velocity magnitude scales the Mach number of the vorticity fields and thus determines the compressibility and nonlinearity of the system.
The acoustic frequency is introduced through the Strouhal number, which compares the convective time scale of the vorticity field with the oscillation period of the flow.
A combination of both parameters determines sound sources with varying amplitudes and frequencies.
The wavelength of the Stokes layer $\lambda_s$ depends on the Reynolds number with a similarity $\lambda_s \sim d/\sqrt{\Rey\, \St}$.
Different slit thickness will change $\Rey$ and $\St$ as well. 
To separate the effects attributed to $\Rey$ and $\St$, we vary the fluid property dynamic viscosity to adjust for varying $\Rey$.

\section{Spectral analysis of the DNS data}
\label{sec:spectral}

Previous methods typically quantify overall dissipation in an integral sense, similar to power absorption coefficients~\citep{tam2001numerical}, and thus often overlook both the spatial and spectral structure of acoustic energy dissipation and its associated energy budget.
To address this, we use spectral proper orthogonal decomposition (SPOD) to decompose the acoustic-induced flow-field fluctuations into optimally energy-ranked coherent spectral flow structures.
This analysis enables separate, mode-by-mode investigation of the kinetic energy (KE) and viscous loss (VL) contributions at each frequency, together with their corresponding coherent spatial structures.
Spatial or spectral integration is then used to recover the overall integral behavior, enabling a better understanding of the acoustic energy dissipation mechanism.

\subsection{Spectral Proper Orthogonal Decomposition}

Consider a time series consisting of $N_t$ snapshots.
Each snapshot is represented by an observable state vector $\vec{\mathbf{y}}\in\mathbb{C}^N$, where $N$ is the dimension of the data.
In line with common practice in the analysis of compressible turbulence~\citep{kida1990energy,wang2013cascade}, the observable is the fluctuating density-weighted velocity field, defined as
\begin{align}
  \vec{\mathbf{y}}(\vec{\mathbf{x}};t) = 
    \sqrt{\vec{\mathbf{\beta}}(\vec{\mathbf{x}};t)} \,
    \vec{\mathbf{u}}(\vec{\mathbf{x}};t),
\end{align}
where $\beta = \rho/\rho_0$ is the dimensionless density ratio, quantifying deviations from the nominal reference density, $\rho_0$.
With this definition, the total kinetic energy within the computational domain can be expressed using the energy norm $\|\cdot\|_y$, induced by the inner product, as 
\begin{align}\label{eqn: inner_product}
    K \equiv \int_\Omega \frac{1}{2}\beta(\vb*{x}) \vb*u^{\mathrm{H}}(\vb*{x})\vb*u(\vb*{x})\,\dd \vb*{x} \approx 
        \frac{1}{2} \vec{\mathbf{y}}^\mathrm{H}\mathbf{W}_y \vec{\mathbf{y}} =\frac{1}{2}
        \langle\vec{\mathbf{y}},\vec{\mathbf{y}}\rangle_y \equiv  \frac{1}{2} \|\vec{\mathbf{y}}\|_y^2.
\end{align}
Here, $\mathbf{W}_y$ is a Hermitian positive-definite weight matrix for the discrete approximation of the volume integral, $\sum_{\Omega}( \, \cdot \, )\Delta_x\Delta_y$, and $( \, \cdot \, )^{\mathrm{H}}$ denotes the Hermitian transpose.
The density ratio is introduced solely as a weighting function for the velocity field.
For nearly incompressible flows, as in the present configuration, density variations are weak, and their weighting has only a minor effect ($\beta\approx1$).
Nevertheless, we retain this formulation to account for the contributions of both velocity and density fluctuations to the dissipation.

We seek spatiotemporal structures that optimally represent the spectral content of $\vec{\mathbf{y}}$, expressed through its Fourier expansion,
\begin{align}
    \vec{\mathbf{y}}(t) = \sum_{n=-\infty}^\infty \widehat{\vec{\mathbf{y}}}_n \rme^{\rmi 2\pi f_n t} .
\end{align}
Assuming ergodicity in statistically stationary flows, we apply the Welch's method~\citep{welch1967use} to obtain convergent estimates of the spectral densities.
Specifically, the time series is divided into $N_\mathrm{blk}$ blocks, each containing $N_\mathrm{DFT}$ consecutive snapshots separated by a constant time step $\Delta t$.
Adjacent blocks may overlap by $N_\mathrm{ovlp}$ snapshots.
Guidance on the appropriate selection of these parameters is provided in~\citet{schmidt2020guide}.
In practice, the $m$th block, for $m=1,2,\dots,N_\mathrm{blk}$, is constructed as
\begin{align}
    \vec{\mathbf{y}}^{(m)} &=   \left[
    \begin{matrix}
        \vec{\mathbf{y}}_{1}^{(m)}& \vec{\mathbf{y}}_{2}^{(m)} &\cdots & \vec{\mathbf{y}}_{N_f}^{(m)}
    \end{matrix}\right]  \in \mathbb{C}^{N\times N_\mathrm{DFT}},
\end{align}
and its discrete Fourier transform (DFT) along the temporal domain is denoted by
\begin{align}
    \widehat{{\mathbf{Y}}}^{(m)} &= \left[
    \begin{matrix}
      \widehat{\vec{\mathbf{y}}}_{1}^{(m)}& \widehat{\vec{\mathbf{y}}}_{2}^{(m)} &\cdots & \widehat{\vec{\mathbf{y}}}_{N_f}^{(m)}
    \end{matrix}\right].
\end{align}
Here, only the non-negative Fourier components are collected due to the reality of the data, yielding $N_{f}=\lfloor N_\mathrm{DFT}/2 \rfloor+1$ frequency components.
For each frequency index $n=1,2,\dots,N_f$, the ensemble of $N_\mathrm{blk}$ Fourier realizations at frequency $f_n$ across all data blocks forms a data matrix 
\begin{align}\label{eqn:fourier_recipient}
    \widehat{{\mathbf{Y}}}_n &=   
    \left[
    \begin{matrix}
      \widehat{\vec{\mathbf{y}}}_{n}^{(1)}& \widehat{\vec{\mathbf{y}}}_{n}^{(2)} &\cdots & \widehat{\vec{\mathbf{y}}}_{n}^{(N_{\text{blk}})}
    \end{matrix}\right].
\end{align}
The corresponding cross-spectral density matrix is
\begin{align}
    \mathbf{S}_n \equiv \frac{1}{N_{\rm blk}}\widehat{{\mathbf{Y}}}_n\widehat{{\mathbf{Y}}}_n^{\mathrm{H}}.
\end{align}
The eigenvectors, $\widehat{\mathbf{\Xi}}_{n}$, and the eigenvalues, $\mathbf{\Lambda}_n$, of the eigenvalue problem
\begin{align}\label{spod_cros}
    \mathbf{S}_n \mathbf{W}_y \widehat{\mathbf{\Xi}}_{n}  = \mathbf{\Lambda}_n \widehat{\mathbf{\Xi}}_{n},
\end{align}
give the SPOD modes, which are the columns of $\widehat{\mathbf{\Xi}}_{n}$ and their corresponding modal energy, respectively.
The eigenvalues are sorted in decreasing order of energy (by singular value) as
\begin{gather}\label{spod_eigen}
    \lambda_n^{(1)}\geq\lambda_n^{(2)}\geq\cdots\geq \lambda_n^{(N_\mathrm{blk})}.
\end{gather}
At the same frequency, the SPOD modes are mutually orthogonal as
\begin{gather}\label{eqn:orthogonality}
    \langle \smash{\widehat{\vec{\bm{\mathrm{\xi}}}}}_n^{(k_1)}, \smash{\widehat{\vec{\bm{\mathrm{\xi}}}}}_n^{(k_2)}\rangle_y = 
        \delta_{k_1,k_2},   
\end{gather}
and are optimal for modal energy under the norm induced by equation~\cref{eqn: inner_product}.
Here, $\delta_{\cdot,\cdot}$ is the Kronecker delta function, and the indices $k_1,k_2 =1,2,\cdots,N_\mathrm{blk}$ refer to the modal ranks.
We refer the reader to \citet{towne2018spectral} for more details of SPOD. 
SPOD provides an optimal decomposition of the flow field into energy-ranked structures as
\begin{align}
    \widehat{\vec{\mathbf{y}}}_n  = \sum_{k=1}^{N_k} a_n^{(k)} \smash{\widehat{\vec{\bm{\mathrm{\xi}}}}}_n^{(k)}, 
    \quad \text{where} \quad
    a_n^{(k)} \equiv \langle \smash{\widehat{\vec{\bm{\mathrm{\xi}}}}}_n^{(k)},\widehat{\vec{\mathbf{y}}}_n\rangle_y,
\end{align}
is the corresponding SPOD expansion coefficient. 
Across different ranks at a given frequency, the SPOD expansion
coefficients are uncorrelated~\citep{nekkanti2021frequency,chu2025stochastic}, that is
\begin{align}\label{eqn:uncorrelatedness}
    \mathrm{E}\left\{  \left({a_n^{(k_1)}}\right)^\mathrm{H} a_n^{(k_2)} \right\} = \lambda_{n}^{(k_1)} \delta_{k_1,k_2}.
\end{align}

\subsection{Representative SPOD eigenvalue spectrum and modes}
\label{subsec:representative_spod}

Herein, spectral analysis is performed within a reduced spatial domain, balancing computational efficiency and accuracy.
For cases with $\St \le 8 \St_{0}$, the analysis is confined to a $28d \times 28d$ region adjacent to the slit, and for $\St \ge 10\St_0$, a $22.4d \times 28d$ region is used.
As shown in \cref{subsec:Database_data}, these domains represent over \SI{99}{\percent} of the total KE and VL, ensuring that the essential flow physics associated with the dissipation mechanism are represented.
The SPOD sampling interval, $\Delta t_\mathrm{spod}$, is set to 2000 simulation time steps for $\St \le 8 \St_{0}$ and 1500 time steps for $\St \ge 10\St_0$. 
SPOD is performed using 640 snapshots ($N_t = 640$), partitioned into nine blocks with \SI{87.5}{\percent} overlap with a block length $N_\mathrm{DFT} = 320$.
A rectangular window is adopted as the acoustic signal is periodic over each analysis segment.
Spectral convergence is confirmed by the large separation between the leading and suboptimal SPOD eigenvalues, indicating accurate identification of dominant coherent structures.

\begin{figure}[ht!]
    \centering
    \includegraphics[]{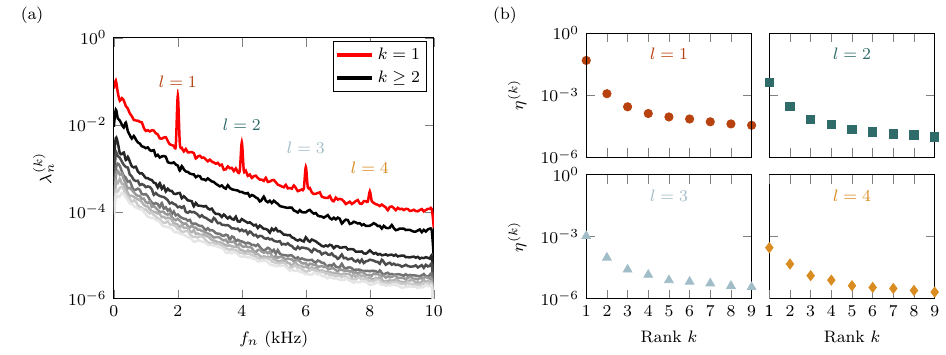}
    \caption{SPOD-based spectral energy analysis for $\mathrm{ISPL} = \SI{150}{\decibel}$, $\St=4\St_0$ and $\Rey=\Rey_0$: 
    (a) eigenvalue spectra, $\lambda^{(k)}_n$; and 
    (b) energy associated with each rank, $\eta^{(k)}$, at four harmonic frequencies ($l=1$--$4$ defined in \cref{eqn:harmonics}).}
    \label{spod_eigenvalue_2kHz} 
\end{figure}

As an example, \cref{spod_eigenvalue_2kHz}~(a) shows the SPOD eigenvalue spectrum for the case of \SI{150}{\decibel} with a Reynolds number of $\Rey_0$, where the imposed acoustic frequency $f_s=\SI{2}{\kilo\hertz}$ corresponds to a Strouhal number of $\St=4\St_0$.
The SPOD eigenvalues of the leading rank exhibit four tonal peaks of decreasing magnitude at integer multiples of $f_s$, implying a harmonic response.
We introduce the harmonic index
\begin{align}\label{eqn:harmonics}
    l \equiv f_n/f_s,
\end{align}
such that the fundamental frequency corresponds to $l=1$ and higher-order harmonics are given by $l=2,3,\cdots,\lfloor (1/(2\Delta t_{\rm spod}  f_s)\rfloor$, up to the Nyquist limit.
These spectral peaks indicate the presence of coherent vortex shedding triggered by the imposed acoustic excitation~\citep{tam2001numerical,tam2005computational}.
The remaining eigenvalues form a broadband spectrum that decreases monotonically in value.
This behavior is characteristic of mixed broadband--tonal flows, whose complex dynamics combine both stochastic and deterministic features.

\Cref{spod_eigenvalue_2kHz}~(b) shows the distribution of energy across the mode ranks for the four tonal peaks.
The magnitude of the leading SPOD rank is at least one order of magnitude larger than that of the suboptimal ranks.
This pronounced separation reflects the distinct low-rank dynamics of the flow field.
As later shown in \cref{spod_eigenvalue}, these trends are qualitatively consistent across all parameter combinations listed in \cref{table:dns_config}.
We define the contribution of each rank $k$, for $k=1,2,\cdots,N_\mathrm{blk}$, via the energy fraction as
\begin{align} \label{eqn:eigenvalue_normalized}
    \eta^{(k)}= \sum_n \eta_n^{(k)} = \frac{\sum_n \lambda_n^{(k)}}{\sum_n\sum_k \lambda_n^{(k)}},
\end{align}
which measures the portion of modal energy.
Accordingly, we focus on the first three ranks, which represent \SI{95}{\percent}, \SI{3.5}{\percent}, and \SI{1}{\percent} of the total energy, respectively.
Collectively, these three modes account for \SI{99}{\percent} of the energy across the examined parameter combinations.
Contributions from higher-order suboptimal modes are neglected in the spectral analysis due to their minimal energy content.

\begin{figure}[ht!]
    \centering
    \includegraphics[]{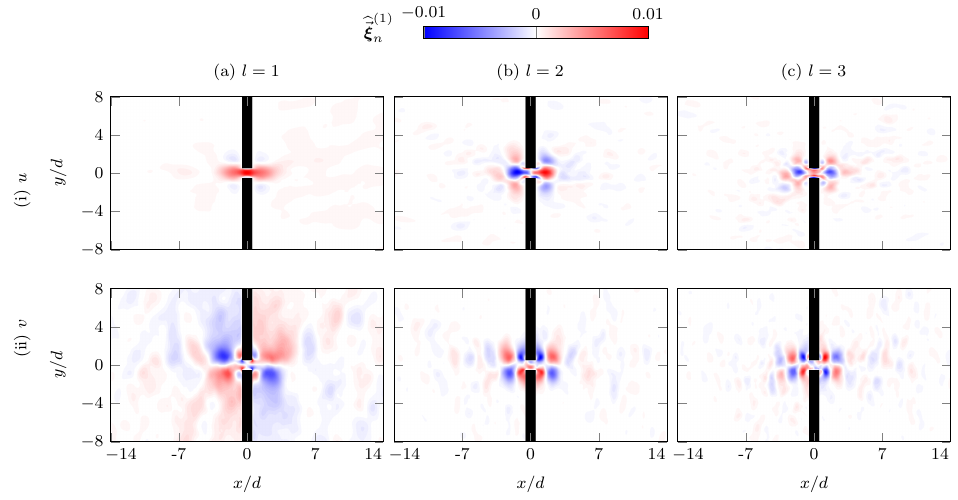}
    \caption{Leading SPOD modes, $\smash{\widehat{\vec{\bm{\mathrm{\xi}}}}}_n^{(1)}$, at $\mathrm{ISPL} = \SI{150}{\decibel}$, $\St=4\St_0$, and $\Rey=\Rey_0$.
    Cases in (a)--(c) correspond to $l=1$, $2$, and $3$.
    Cases in (i) and (ii) correspond to velocities in the $x$- and $y$-coordinate directions, $u$ and $v$.
    }
    \label{spod_modes_2kHz} 
\end{figure}

\Cref{spod_modes_2kHz} shows the leading fundamental, 2nd, and 3rd harmonic SPOD modes for the case examined in \cref{spod_eigenvalue_2kHz}.
The spatial structures are active both upstream and downstream of the slit, reflecting the periodic nature of the acoustically driven flow field.
Across all harmonics, the modes are concentrated near the slit, with pronounced intensities at the four slit corners.
These localized structures highlight strong interactions between the flow and the slit geometry.
In particular, the induced oscillatory boundary layer dynamics enable the conversion of incident acoustic energy into coherent vortical motion.
As expected, the spatial modes exhibit smaller-scale structures as the harmonic order increases.
This trend is consistent with the higher effective excitation frequencies associated with harmonic tones, which induce faster oscillations in the flow field.
The slight asymmetry of the modes about the horizontal centerline, especially in the $y$-direction velocity component, is consistent with observations from previous analyses based on both numerical simulations~\citep{qiang2022aeroacoustic} and experimental measurements~\citep{tang2025phase}.
This symmetry-breaking reflects the inherently irregular nature of the induced vortex dynamics and may develop into full asymmetry as the shed vortices interact and undergo nonlinear amplification.

\subsection{Spectral features across the parameter space}
\label{subsec:parameter_spectrum}

\begin{figure}[ht!]
    \centering
    \includegraphics[]{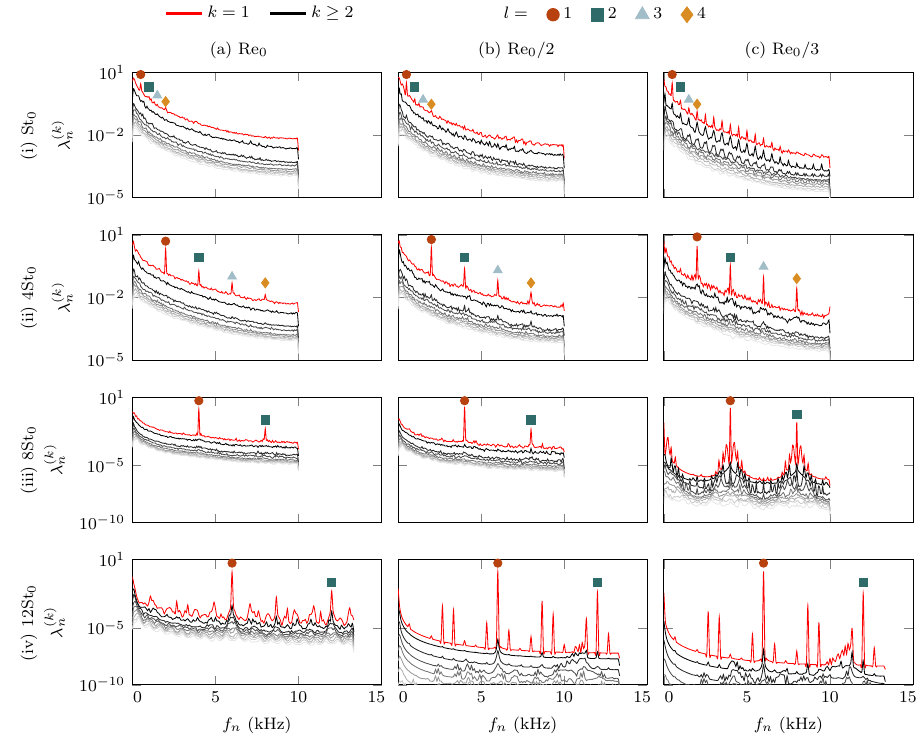}
    \caption{SPOD eigenvalue spectra, $\lambda_n^{(k)}$, at different rank $k$ for different $\St$--$\Rey$ combinations at $\mathrm{ISPL} = \SI{150}{\decibel}$.
     Cases in column (a)--(c) represent $\Rey=\Rey_0$, $\Rey_0/2$, and $\Rey_0/3$. Cases in row (i)--(iv) represent $\St=\St_0$, $4\St_0$, $8\St_0$,and $12\St_0$.
    The fundamental frequency ($l=1$) and its higher-order harmonics ($l \ge 2$) manifest as peaks in the spectrum.
    For example, two tonal peaks arise at $f_n = \SI{4}{\kilo \hertz}$ ($l=1$) and $f_n = \SI{8}{\kilo \hertz}$ ($l=2$) in case (a, iii).
    }
    \label{spod_eigenvalue} 
\end{figure}

To understand the spectral behavior, we compare the SPOD eigenvalue spectra across the parameter space at $\mathrm{ISPL} = \SI{150}{\decibel}$, as shown in \cref{spod_eigenvalue}.
In general, the harmonic peaks become more distinct from the broadband content for smaller $\Rey$.
This trend is consistent with the reduced spectral density at lower $\Rey$, which lowers the broadband energy level and thereby enhances the spectral contrast between coherent harmonic peaks and incoherent broadband components.
In addition, both the leading and suboptimal ranks exhibit more visible harmonic peaks at lower $\Rey$, without a reduction in the separation between their SPOD eigenvalues.
This observation suggests that increased viscosity at smaller $\Rey$ promotes phase locking of large-scale vortical structures across cycles, thereby suppressing broadband fluctuations.
An increase in acoustic excitation frequency also leads to stronger harmonic peaks.
At $\St=12\St_0$, additional peaks appear at non-integer multiples of the forcing frequency and become more pronounced at lower $\Rey$.
We observe similar features at $\St=8\St_0$, but only at the lowest $\Rey$, while other cases exhibit clean broadband-tonal spectra.
These additional peaks suggest more complex nonlinear interactions within the flow field, involving not only vortex shedding harmonics but also interactions with the slit boundaries.
This behavior can be qualitatively observed in the instantaneous fields shown in \cref{instantaneous_vorticity_150db}, where the shed vortices at $\St=12\St_0$ travel along and interact with the $y$-direction boundaries of the slit, rather than propagating in phase with the acoustic wave.

Doubling the characteristic length scale, $d$, in the case where $\St_1 = 4\St_0$ and $\Rey_1 = \Rey_0/2$ results in a new configuration with $\St_2 = 8\St_0$ and $\Rey_2 = \Rey_0$.
Although this transformation does not strictly preserve dynamic similarity, it provides a metric for a parametric trajectory governed by the Womersley number, which is 
\begin{align}\label{eqn: womersley}
    \Wo \equiv \frac{d}{2} \sqrt{\frac{2 \pi f_s}{\nu}}.
\end{align}
The Womersley number, $\Wo$, was introduced to characterize the relative importance of inertial and viscous forces in pulsatile pipe flows~\citep{womersley1955method}, and scales linearly with the characteristic length.

The geometric scaling among $\St$, $\Rey$, and $\Wo$ can be expressed as
\begin{align}\label{eqn: womersley_similarity}
     \Wo \sim \sqrt{\Rey \, \St} = \sqrt{(c \, d/\nu) \, (f_s \, d /c)} = d \sqrt{f_s/ \nu}.
\end{align}
Thus, doubling the slit thickness (i.e., the effective length scale in this study) doubles $\Wo$ and results in a SPOD eigenvalue spectrum with more distinguishable peaks and weaker broadband levels.
A similar trend is observed in the comparison between the cases with
$\St_1 = 4\St_0$ and $\Rey_1 = \Rey_0/3$ and $\St_2 = 12\St_0$ and $\Rey_2 = \Rey_0$, where the latter corresponds to a tripling of the length scale.
Additional cases supporting this observation, not shown in \cref{spod_eigenvalue}, are provided in \cref{spod_eigenvalue_restcases} in \hyperref[appendixA]{Appendix~A}.
This observation is conceptual, and we do not vary $d$ in the DNS.
In practical resonator designs, changes in slit thickness alter $\St$ and $\Rey$ through geometric changes, with an additional dependence on the properties of the background fluid.

\begin{figure}[ht!]
    \centering
    \includegraphics[]{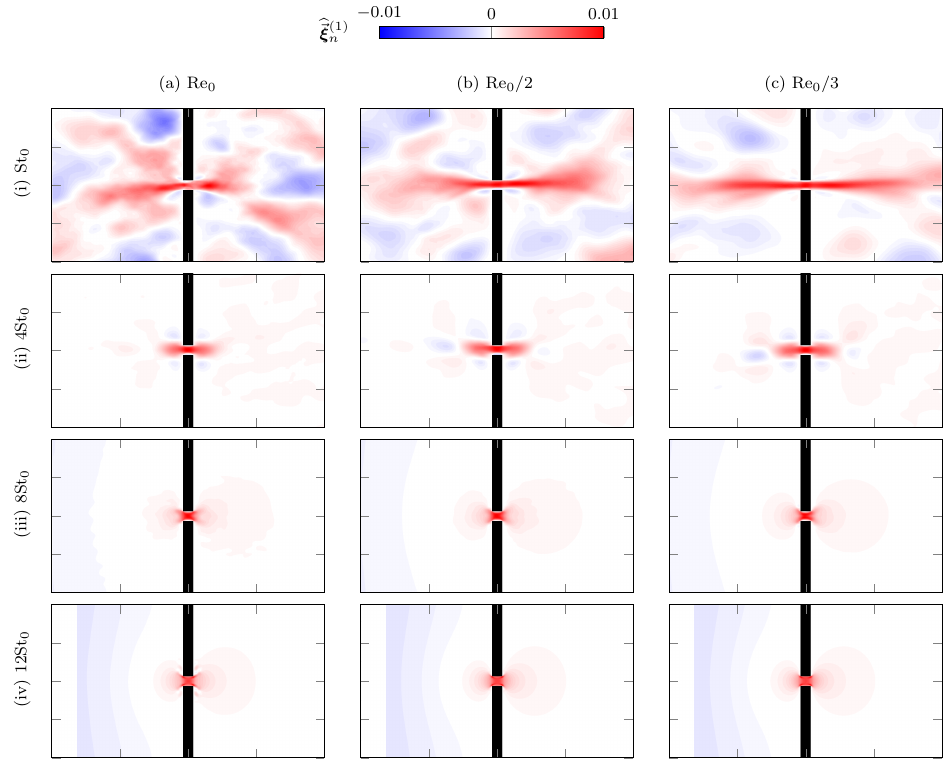}
    \caption{
        Leading SPOD modes, $\smash{\widehat{\vec{\bm{\mathrm{\xi}}}}}_{l=1}^{(1)}$, for different $\St$--$\Rey$ combinations at $\mathrm{ISPL} = \SI{150}{\decibel}$, plotted at their respective fundamental frequencies ($l=1$) as their $x$-directional components, $u$.
        Cases in column (a)--(c) represent $\Rey=\Rey_0$, $\Rey_0/2$, and $\Rey_0/3$.
        Cases in row (i)--(iv) represent $\St=\St_0$, $4\St_0$, $8\St_0$,and $12\St_0$.
    }
    \label{spod_mode1}
\end{figure}

\Cref{spod_mode1} presents the leading $x$-direction SPOD modes at their respective fundamental frequencies, corresponding to the cases shown in \cref{spod_eigenvalue}.
The spatial axes follow the same convention as in \cref{spod_modes_2kHz}, but are omitted here for clarity.
At the lowest acoustic frequency of $\St = \St_{0}$, a decrease in Reynolds number leads to more flattened large-scale structures along the horizontal centerline, while the smaller-scale structures are diminished.
Combined with the corresponding SPOD eigenvalue spectra, where lower-$\Rey$ cases exhibit reduced broadband energy and more pronounced spectral peaks, these observations suggest that stronger viscous effects at this frequency enhance the coherence of vortex shedding. This enhancement occurs while suppressing the remaining flow content.

For larger acoustic frequency, the leading SPOD modes exhibit similarly compact spatial support near the slit, with reduced dependence on $\Rey$.
Specifically, at $\St = 4\St_{0}$, the modes display dumbbell-shaped structures spanning the slit, whereas at $\St = 8\St_{0}$ and $12\St_{0}$, they exhibit X-shaped patterns localized near the slit.
With a virtually doubled slit thickness (i.e., $\Wo$ doubled), the case with $\St = 8\St_0$ and $\Rey = \Rey_0$ exhibits a more confined X-shaped pattern as opposed to the case with $\St = 4\St_0$ and $\Rey = \Rey_0/2$, which appears to be dumbbell-shaped.
\Cref{spod_mode_restcases} in \hyperref[appendixA]{Appendix A} documents the modal structures of the cases not shown in \cref{spod_mode1}.

Taken together, these observations indicate that increasing the acoustic frequency or the slit thickness compresses the dominant vortex shedding structures toward the near-slit region.
At the same time, more confinement can introduce complex interactions near and within the slit.

\begin{figure}[ht!]
    \centering
    \includegraphics[]{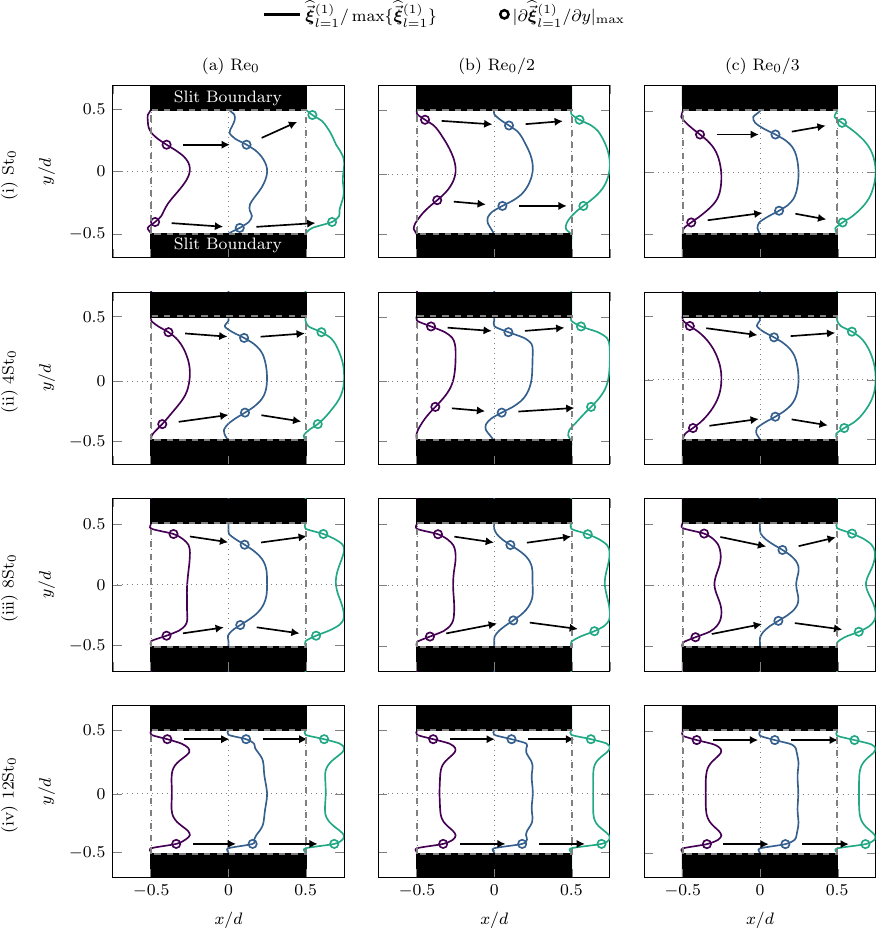}
    \caption{
    Profiles of the $x$-directional components, $u$, of the leading SPOD modes at the fundamental frequency, \raisebox{-0.3ex}{$\smash{\widehat{\vec{\bm{\mathrm{\xi}}}}}_{l=1}^{(1)}/\max\{\smash{\widehat{\vec{\bm{\mathrm{\xi}}}}}_{l=1}^{(1)}\}$}, within the slit for different $\St$--$\Rey$ combinations at $\mathrm{ISPL} = \SI{150}{\decibel}$.
    The maximum mode gradient in $y$ direction on on the top and bottom halves of each profile, \smash[t]{\raisebox{-0.3ex}{$|\partial \smash{\widehat{\vec{\bm{\mathrm{\xi}}}}}_{l=1}^{(1)}/\partial y|_\mathrm{max}$}}, is marked as a hollow circle, which is connected by arrows across all the profiles.
    Cases in column (a)--(c) correspond to $\Rey=\Rey_0$, $\Rey_0/2$, and $\Rey_0/3$.
    Cases in row (i)--(iv) correspond to $\St=\St_0$, $4\St_0$, $8\St_0$, and $12\St_0$.
    }
    \label{spod_BL} 
\end{figure}

To have a detailed understanding of mode structure inside the slit, we show the normalized velocity profiles at the inlet ($x=-d/2$), center ($x=0$), and outlet ($x=d/2$) of the slit, respectively, in \cref{spod_BL}.
For most $\St$--$\Rey$ combinations, the inlet and outlet profiles are nearly identical, indicating weak $x$-direction variation of the fundamental mode between the inlet and outlet.
We calculated the mode gradient in the $y$ direction using first-order finite-difference approximation. We highlighted two points on each profile, where the maximum mode gradients are present for both the top and bottom halves of the profile.
These points have implications on the distribution of the VL field, which will be discussed in \cref{subsec:dissipation_spectrum}.
At the lower $\St \le 4\St_0$ and higher $\Rey \ge \Rey_0/2$, the profiles are asymmetric about the horizontal centerline.
In this regime, the oscillatory motion produces a thicker boundary layer, and we observe mild boundary-layer separation near the slit walls, with wall shear sign changes, consistent with the larger-scale mode structures seen in \cref{spod_mode1}.
In contrast, at $\St > 4\St_0$ and $\Rey < \Rey_0/2$, the profile is similar to the Womersley flow profile with a Womersley number $\Wo \ge 16$~\citep{womersley1955method}, exhibiting symmetrical structure according to the horizontal centerline.
The increased acoustic frequency yields a thinner boundary layer without apparent separation.

\subsection{Dissipation Mechanism}
\label{subsec:dissipation_spectrum}

By decomposing the flow field into optimally energy-ranked SPOD modes, we analyze the total dissipation mechanism for each spectral component as vortical KE of the induced flow field and VL contributions.
For each frequency $f_n$, we express the velocity and stress tensor as linear combinations of the SPOD modes, $\smash[t]{\widehat{\vec{\bm{\mathrm{\xi}}}}}_n^{(k)}$, and corresponding expansion coefficients, $a_n$, as
\begin{align}
    \widehat{\vec{\mathbf{y}}}_n(\vec{\mathbf{x}}) \approx \sum_{k=1}^{3} a_n^{(k)}\smash{\widehat{\vec{\bm{\mathrm{\xi}}}}}_n^{(k)}(\vec{\mathbf{x}}),
    \quad \text{and} \quad 
    \widehat{\vec{\mathbf{T}}}_n(\vec{\mathbf{x}})  \approx \sum_{k=1}^{3} a_n^{(k)}\smash{\widehat{\vec{\bm{\mathrm{T}}}}}_n^{(k)}(\vec{\mathbf{x}}),
\end{align}
where 
\begin{gather}
    \smash{\widehat{\vec{\bm{\mathrm{T}}}}}_n^{(k)}(\vec{\mathbf{x}}) \equiv\frac{1}{\overline{\beta} \Rey} \left[\grad \smash{\widehat{\vec{\bm{\mathrm{\xi}}}}}_n^{(k)}+\left(\grad \smash{\widehat{\vec{\bm{\mathrm{\xi}}}}}_n^{(k)}\right)^{\mathrm{H}} -\frac{2}{3}\left(\grad\vdot \smash{\widehat{\vec{\bm{\mathrm{\xi}}}}}_n^{(k)}\right)\vec{\mathbf{I}}\right](\vec{\mathbf{x}})
\end{gather}
is the rank-$k$ stress-tensor mode.

To enable a direct, dimensionally consistent comparison between the spectral KE and VL components, we weight the KE spectrum by the corresponding angular frequency, \(\omega_n=2\pi f_n\).
This process converts the spectral KE density (energy per frequency) into an energy-rate scale (power per frequency) with units consistent with the spectral VL density.
Frequency weighting is introduced for dimensional consistency and comparison, while a budget interpretation requires applying the same scaling consistently to all terms of the spectral KE equation.
The frequency-weighted spectral KE field can then be expressed as
\begin{align}
    \widehat{\vec{{K}}}_n(\vec{\mathbf{x}}) =  \frac{1}{2}\mathrm{E}\left\{\left[\omega_n {\widehat{\vec{\mathbf{y}}}_n}^\mathrm{H}{\widehat{\vec{\mathbf{y}}}_n} \right](\vec{\mathbf{x}})\right\}
     & \approx \pi f_n \mathrm{E}\left\{\left[ {\left(\sum_{k=1}^{3}a_n^{(k)}\smash{\widehat{\vec{\bm{\mathrm{\xi}}}}}_n^{(k)}\right)}^\mathrm{H}\left(\sum_{k=1}^{3}a_n^{(k)}\smash{\widehat{\vec{\bm{\mathrm{\xi}}}}}_n^{(k)}\right) \right](\vec{\mathbf{x}})\right\}\\
    &= \pi f_n  \mathrm{E}\left\{ \sum_{k_1=1}^{3} \sum_{k_2=1}^{3}\left[  \left({a_n^{(k_1)}}\right)^\mathrm{H} a_n^{(k_2)} \left({\smash{\widehat{\vec{\bm{\mathrm{\xi}}}}}_n^{(k_1)}}\right)^\mathrm{H} \smash{\widehat{\vec{\bm{\mathrm{\xi}}}}}_n^{(k_2)} \right](\vec{\mathbf{x}})\right\} \label{eqn:Kn_1}\\
   & = \pi f_n \sum_{k=1}^{3} \left[ \lambda_{n}^{(k)} \left({\smash{\widehat{\vec{\bm{\mathrm{\xi}}}}}_n^{(k)}}\right)^\mathrm{H} \smash{\widehat{\vec{\bm{\mathrm{\xi}}}}}_n^{(k)} \right](\vec{\mathbf{x}}) = \sum_{k=1}^{3}  \smash{\widehat{\vec{\bm{K}}}}_n^{(k)} (\vec{\mathbf{x}})\label{eqn:Kn_2}.
\end{align}
The simplification from equation \cref{eqn:Kn_1} to equation \cref{eqn:Kn_2} follows from the mutual uncorrelatedness of the SPOD coefficients shown in equation~\cref{eqn:uncorrelatedness}.
Following the canonical treatment of VL~\citep{landau1987fluid}, the total spectral VL can be decomposed into shear (deviatoric) and dilatational (compressible) contributions by separating the terms of the viscous stress tensor in equation~\cref{eq:viscous_stress} into its symmetric and divergence-related parts.
In particular, we focus on the shear-driven contribution associated with the symmetric strain-rate tensor,
\begin{align}
        \widehat{\vec{{D}}}_n(\vec{\mathbf{x}})
        \approx\mathrm{E}\left\{\left[\widehat{\vec{\mathbf{T}}}_n:\grad {\widehat{\vec{\mathbf{y}}}_n} \right](\vec{\mathbf{x}})\right\} 
        &\approx    \mathrm{E}\left\{\left[ \left( \sum_{k=1}^{3} a_n^{(k)}\smash{\widehat{\vec{\bm{\mathrm{T}}}}}_n^{(k)} \right):\grad {\left(\sum_{k=1}^{3}a_n^{(k)}\smash{\widehat{\vec{\bm{\mathrm{\xi}}}}}_n^{(k)} \right)} \right](\vec{\mathbf{x}})\right\}\\
        & = \sum_{k=1}^{3} \lambda_n^{(k)}\left[ \smash{\widehat{\vec{\bm{\mathrm{T}}}}}_n^{(k)}:\grad \smash{\widehat{\vec{\bm{\mathrm{\xi}}}}}_n^{(k)} \right](\vec{\mathbf{x}}) = \sum_{k=1}^{3} \smash{\widehat{\vec{D}}}_n^{(k)} (\vec{\mathbf{x}}).
\end{align}
which represents vortex-related dissipation.
Here $\smash{\widehat{\raisebox{-0.5ex}{$\vec{D}$}}}_n^{(k)}$
and $\smash{\widehat{\raisebox{-0.5ex}{$\vec{D}$}}}_n$ are effectively real with negligible imaginary components.
The compressible term depends on the dilatation, $\grad\cdot \hat{\vec{\mathbf{u}}}$, and associated density gradient, $\grad \beta$, which together represent volumetric and pressure-related damping.
As shown in \cref{spod_compressibility} in \hyperref[appendixC]{Appendix C}, the compressible term contributes less than \SI{1}{\percent} to the total, and can be reasonably neglected in the present analysis.

The rank-$k$ spectral KE and VL fields take the forms of
\begin{gather}
    \smash{\widehat{\vec{K}}}_n^{(k)} (\vec{\mathbf{x}}) = \pi f_n  \lambda_{n}^{(k)} \left[  \left({\smash{\widehat{\vec{\bm{\mathrm{\xi}}}}}_n^{(k)}}\right)^\mathrm{H} \smash{\widehat{\vec{\bm{\mathrm{\xi}}}}}_n^{(k)} \right](\vec{\mathbf{x}})
    \quad \text{and} \quad
    \smash{\widehat{\vec{D}}}_n^{(k)}(\vec{\mathbf{x}}) = \lambda_{n}^{(k)}
       \left[ \smash{\widehat{\vec{\bm{\mathrm{T}}}}}_n^{(k)}:\grad \smash{\widehat{\vec{\bm{\mathrm{\xi}}}}}_n^{(k)} \right](\vec{\mathbf{x}}),
\end{gather}
respectively.
Their respective integral measures within the computational domain are
\begin{gather}
    \widehat{{{K}}}_n^{(k)} = 
    \sum_{\Omega}\smash{\widehat{\vec{K}}}_n^{(k)}(\vec{\mathbf{x}}) \Delta_x\Delta_y = \pi f_n \lambda_n^{(k)}, \quad \text{and} \quad 
    \widehat{{{D}}}_n^{(k)} = \sum_{\Omega}\smash{\widehat{\vec{D}}}_n^{(k)}(\vec{\mathbf{x}}) \Delta_x\Delta_y,
\end{gather}
where the normality of SPOD modes in equation~\cref{eqn:orthogonality} is used.
Summing the contributions from the first three ranks yields
\begin{gather}
    \widehat{{{K}}}_n = 
    \sum_{\Omega}\widehat{\vec{{K}}}_n(\vec{\mathbf{x}}) \Delta_x\Delta_y = \pi f_n \left(\lambda_n^{(1)}+\lambda_n^{(2)}+\lambda_n^{(3)}\right), \quad \text{and} \quad 
    \widehat{{{D}}}_n = \sum_{\Omega} \widehat{\vec{{D}}}_n(\vec{\mathbf{x}}) \Delta_x\Delta_y.
\end{gather}
We approximate the integration over the discrete frequency $f_n$ up to the Nyquist frequency using the summation operator:
\begin{gather} \label{eqn:f_operator}
    \mathcal{S}( \, \cdot \, )=\sum_{f_n\in \left[0, 1/(2\Delta t_{\rm spod})\right]}( \, \cdot \, )\,\Delta f_n,
\end{gather}
where $\Delta f_n=1/(N_\mathrm{DFT}\Delta t_{\rm spod})$ is the frequency-bin width.
In practice, the sum is evaluated using the trapezoidal rule.
The rank-$k$ contributions, obtained by integrating over the frequency domain, are defined as
\begin{align}
    \widehat{\mathcal{K}}^{(k)} = \mathcal{S} (\widehat{{K}}_n^{(k)}), 
    \quad \text{and} \quad   
    \widehat{\mathcal{D}}^{(k)} =  \mathcal{S} (\widehat{{D}}_n^{(k)}),
\end{align}
respectively.
Likewise, the corresponding total (all-rank) contributions are
\begin{align}
    {\mathcal{K}} = \mathcal{S} (\widehat{{K}}_n), 
    \quad \text{and} \quad   
    {\mathcal{D}} =  \mathcal{S} (\widehat{{D}}_n).
\end{align}
From Parseval's theorem, the spectral integral of the viscous-loss density, $\mathcal{D}$, recovers the mean irreversible dissipation power, $\overline{D}$.
The integrated frequency-weighted KE is a power-like spectral moment satisfying $\mathcal{K}=\overline{K}\overline{\omega}_k$, where $\overline{K}$ is the mean KE and $\overline{\omega}_k$ is the energy-weighted characteristic angular frequency.
In general, $\mathcal{K}$ provides a convenient rate scale for comparison with $\mathcal{D}$ but does not, by itself, imply closure of the spectral KE budget.
In the present study, the KE spectrum is strongly concentrated around the vortex-shedding frequency such that $\overline{\omega}_k$ is dominated by that peak, making $\mathcal{K}$ an effective single-timescale measure of the KE dynamics.
This interpretation is consistent with the metric used by \citet{tam2001numerical}, in which a cycle-based energy conversion is considered by quantifying the KE of large-scale shedding vortices per acoustic period.

\begin{figure}[ht!]
    \centering
    \includegraphics[]{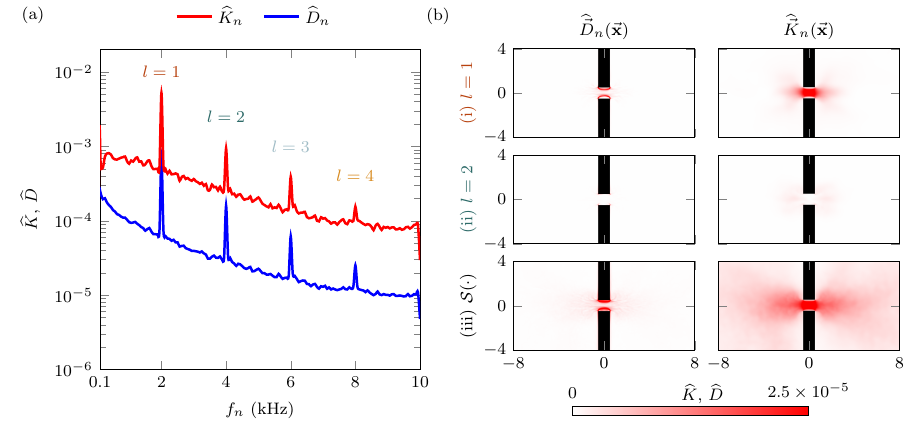}
    \caption{
        Acoustic energy dissipation at $\mathrm{ISPL} = \SI{150}{\decibel}$, $\St = 4\St_0$ and $\Rey = \Rey_0$: 
        (a) energy dissipation spectra; and (b) spectral VL and frequency-weighted KE fields at harmonic indices $l=1$ (i) and $l=2$ (ii), together with the field as $\mathcal{S}\left(\smash{\widehat{\raisebox{-0.5ex}{$\vec{D}$}}_n}\right)(\vec{\mathbf{x}})$ for VL and $\mathcal{S}\left(\smash{\widehat{\raisebox{-0.5ex}{$\vec{K}$}}_n}\right)(\vec{\mathbf{x}})$ for KE contributions of all spectral components (iii).}
    \label{spod_DKE_2kHz} 
\end{figure}


Consistent with this representative case of $\mathrm{ISPL} = \SI{150}{\decibel}$, $\St=4\St_0$ analyzed in \cref{spod_eigenvalue_2kHz}, 
\cref{spod_DKE_2kHz} presents the corresponding spectral KE and VL spectra, computed from the contributions of the first three leading SPOD modes.
Within the analysis domain $\Omega$, incident acoustic energy is primarily damped by conversion into vortical kinetic energy of the induced flow fluctuations and irreversible viscous attenuation~\citep{tam2001numerical}.
We quantify these contributions using the spectral KE and VL components and define the total mode-frequency-resolved contribution as
\begin{align}
\smash{\widehat{\vec{P}}}_n^{(k)}(\vec{\mathbf{x}})= \smash{\widehat{\vec{K}}}_n^{(k)}(\vec{\mathbf{x}})+\smash{\widehat{\vec{D}}}_n^{(k)}(\vec{\mathbf{x}}), \quad 
    \widehat{P}^{(k)}_n \equiv \widehat{K}^{(k)}_n + \widehat{D}^{(k)}_n \quad \text{and} \quad {\mathcal{P}} \equiv {\mathcal{K}} + {\mathcal{D}}.
\end{align}
Here, $\smash{\widehat{\raisebox{-0.5ex}{$\vec{P}$}}}_n^{(k)}$ provides an estimate of the total spectral density of acoustic energy dissipation for mode $k$, combining energy stored in coherent vortical motion and energy irreversibly converted to heat through viscosity.
In this representative case, the $\widehat{D}_n$ is roughly \SI{20}{\percent} of $\widehat{K}_n$, indicating that most of the incident acoustic energy loss results from conversion into vortical kinetic energy.
In contrast, viscous losses still make a non-negligible contribution.

\Cref{spod_DKE_2kHz}~(b) shows the spectral KE and VL fields associated with the fundamental frequency, 2nd harmonic frequency, and spectral integration, respectively.
At the fundamental frequency, both the spectral VL and KE fields show dominant structures, with the VL field concentrated within the slit and the KE field extending slightly outside the slit opening.
At the second harmonic, the spectral VL field is localized near the slit corners.
Due to the higher frequency, the associated vorticity exhibits reduced coherence.
The vorticity cannot be convected as easily into the central region of the slit, resulting in concentrated dissipation at the slit periphery.
This localization is consistent with the reduced propagation distance of high-frequency vortical structures.
The spectral KE field at the second harmonic is primarily located outside the slit, with its magnitude approximately \SI{25}{\percent} of that at the fundamental frequency.
The limited spatial extent and energy content of this mode suggest that the kinetic energy is dissipated before forming large-scale coherent structures, which aligns with the localized dissipation observed in the spectral VL field.
The overall VL field, $\mathcal{S}\left(\smash{\widehat{\raisebox{-0.5ex}{$\vec{D}$}}_n}\right)(\vec{\mathbf{x}})$, remains confined to the slit region, reinforcing the role of boundary-layer interactions as the dominant mechanism of viscous energy loss.
In contrast, the overall KE field, $\mathcal{S}\left(\smash{\widehat{\raisebox{-0.5ex}{$\vec{K}$}}_n}\right)(\vec{\mathbf{x}})$, is more broadly distributed across the domain, with significant upstream and downstream structures.

\begin{figure}[ht!]
    \centering
    \includegraphics[]{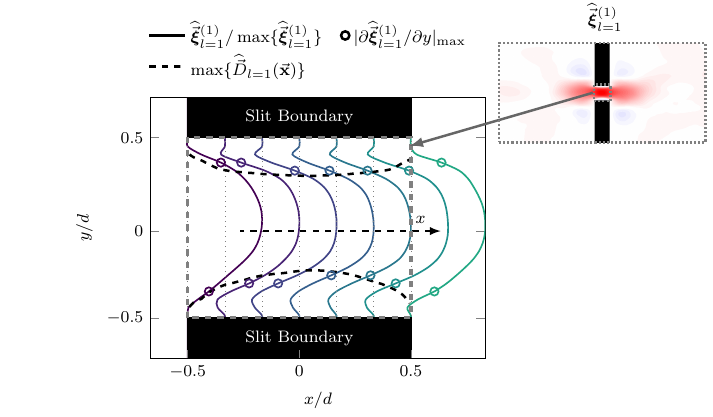}
    \caption{
    Profiles of the $x$-directional components, $u$, of the leading SPOD modes at the fundamental frequency, \raisebox{-0.3ex}{$\smash{\widehat{\vec{\bm{\mathrm{\xi}}}}}_{l=1}^{(1)}/\max\{\smash{\widehat{\vec{\bm{\mathrm{\xi}}}}}_{l=1}^{(1)}\}$}, within the slit for $\mathrm{ISPL} = \SI{150}{\decibel}$, $\St=4\St_0$,and $\Rey=\Rey_0$.
    The maximum mode gradient in $y$ direction on the top and bottom halves of each profile, \smash[t]{\raisebox{-0.3ex}{$|\partial \smash{\widehat{\vec{\bm{\mathrm{\xi}}}}}_{l=1}^{(1)}/\partial y|_\mathrm{max}$}}, is marked as a hollow circle.
    The maximum mode gradient aligns the maximum VL at the fundamental frequency observed in \cref{spod_DKE_2kHz}, using $\smash{\widehat{\raisebox{-0.75ex}{$\vec{D}$}}_{l=1}}(\vec{\mathbf{x}})$.
    }
    \label{spod_BL_2kHz} 
\end{figure}

To confirm the role of the boundary layer in viscous dissipation, \cref{spod_BL_2kHz} shows the normalized $x$-direction velocity profiles of the fundamental mode within the slit.
These profiles are used to assess the spatial alignment between regions of maximum velocity gradients and maximum VL.
The profiles exhibit a Poiseuille-like core in the central region of the slit, primarily driven by the pressure gradient between the upstream and downstream sides of the slit.
This pressure difference also reflects the energy loss of the acoustic waves as they propagate through the slit.
Near the walls, the profiles show evidence of flow separation, indicative of an adverse pressure gradient and consistent with the strong boundary-layer dissipation observed in the spectral VL field.
The steepest velocity gradients occur within these boundary-layer separation zones, where the interaction between the oscillatory motion and the no-slip boundary condition generates intense shear.
This separation results in a thicker and more asymmetric dissipation field, consistent with previously discussed modal asymmetries.
The spectral VL field is concentrated within the oscillatory boundary layer, aligning with regions where the $y$-direction gradient of $x$-direction velocity, $\partial u/\partial y$, is maximum.
This alignment confirms that unsteady boundary-layer dynamics are responsible for viscous energy losses.

\begin{figure}[ht!]
    \centering
    \includegraphics[]{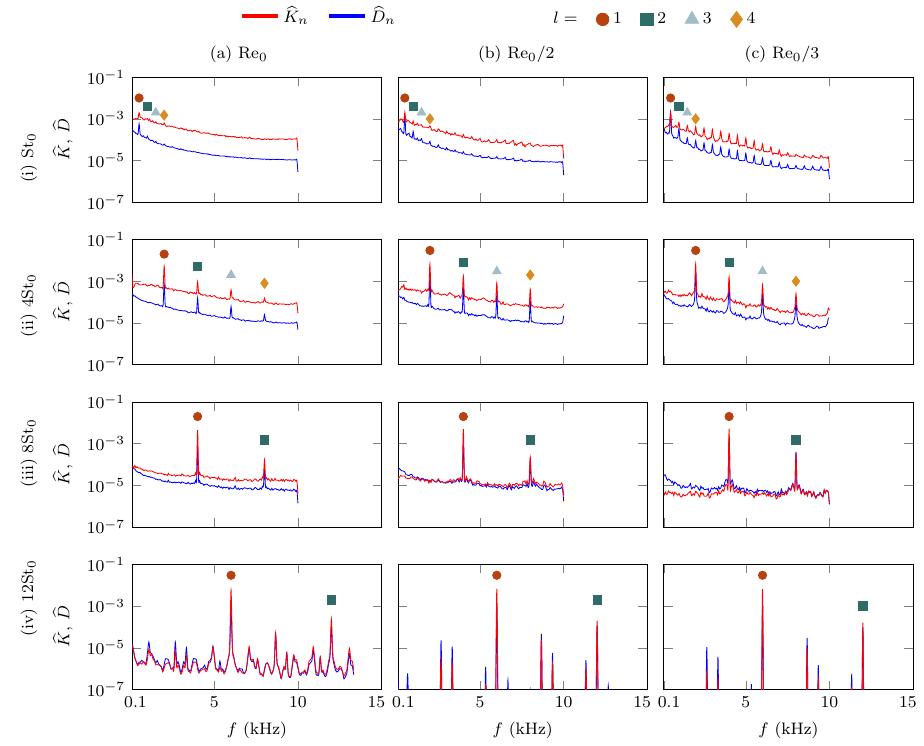}
    \caption{
       Frequency-weighted KE and VL spectra for different $\St$--$\Rey$ combinations at $\mathrm{ISPL} = \SI{150}{\decibel}$.
        The fundamental frequencies $(l=1)$ and their higher-order harmonics $(l\geq 2)$ are manifested as distinct peaks in the spectra.
        Cases in column (a)--(c) represent $\Rey=\Rey_0$, $\Rey_0/2$, and $\Rey_0/3$. Cases in row (i)--(iv) represent $\St=\St_0$, $4\St_0$, $8\St_0$,and $12\St_0$.
    }
    \label{spod_DKE_spectra} 
\end{figure}

\Cref{spod_DKE_spectra} displays the KE and VL spectra, quantified via spatial integration, across the parameter space, illustrating the dependence of energy production and viscous dissipation on $\St$, and $\Rey$ at a fixed input sound pressure level of $\mathrm{ISPL} = \SI{150}{\decibel}$.
As $\Rey$ decreases, the viscous loss in the broadband region becomes comparable to the KE rate, and the difference in peak magnitude between KE and VL also diminishes.
At higher $\St$, the VL spectrum even exceeds the KE spectrum at certain frequencies.
This trend is consistent with the inverse proportional relation $\widehat{D}_n \propto 1/\Rey$, indicating that lower Reynolds numbers suppress irregular activity and reduce the relative contribution of broadband energy dissipation.


At higher $\St$, the spectra are dominated by tonal peaks.
When $\St/\St_0>4$, the peak amplitudes in both KE and VL spectra exceed the broadband levels by at least two orders of magnitude, suggesting minimal presence of irregular fluctuations.
In the extreme case of $\Rey/\Rey_0 = 1/3$ and $\St/\St_0 = 12$, the tonal peak surpasses the broadband level by over six orders of magnitude.
Conversely, the opposite extreme case of $\Rey/\Rey_0 = 1$ and $\St/\St_0 = 1$ shows a tonal-to-broadband contrast of less than one order of magnitude, indicating stronger irregular activity.
These observations are consistent with the SPOD eigenvalue spectra, further confirming that increasing the slit thickness enhances viscous dissipation and leads to more pronounced tonal features.
Additional KE and VL spectra at $\mathrm{ISPL} = \SI{150}{\decibel}$ have similar features and are included in \cref{spod_DKE_restcases} in \hyperref[appendixA]{Appendix A}.

\begin{figure}[ht!]
    \centering
    \includegraphics[]{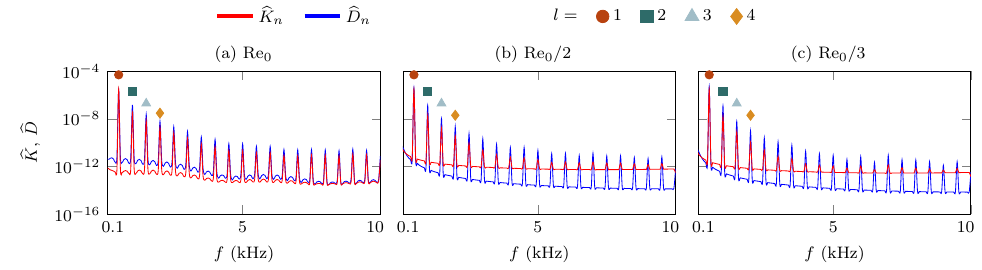}
    \caption{
        Frequency-weighted KE and VL spectra for different $\Rey$ at $\St = \St_0$ and $\mathrm{ISPL} = \SI{120}{\decibel}$ .
        The fundamental frequencies $(l=1)$ and their higher-order harmonics $(l\geq 2)$ are manifested as distinct peaks in the spectra.
        Cases in column (a)--(c) represent $\Rey=\Rey_0$, $\Rey_0/2$, and $\Rey_0/3$.
    }
    \label{spod_DKE_120dB} 
\end{figure}

For comparison, \cref{spod_DKE_120dB} displays the KE and VL spectra at $\St/\St_0 = 1$ and a lower input sound pressure level of $\mathrm{ISPL} = \SI{120}{\decibel}$, across varying $\Rey$.
At this reduced acoustic forcing, the tonal peaks remain prominent, exceeding the broadband levels by at least four orders of magnitude in both KE and VL spectra.
However, the overall energy levels are significantly diminished.
The entire spectrum is approximately six orders of magnitude lower than that observed at $\mathrm{ISPL}=\SI{150}{\decibel}$, reflecting the strong dependence of both KE and viscous dissipation on the ISPL.
Dissipation spectra for additional cases are provided in \cref{spod_DKE_spectra_120dB_full} in \hyperref[appendixB]{Appendix B}.

\begin{figure}[!ht]
    \centering
    \includegraphics[]{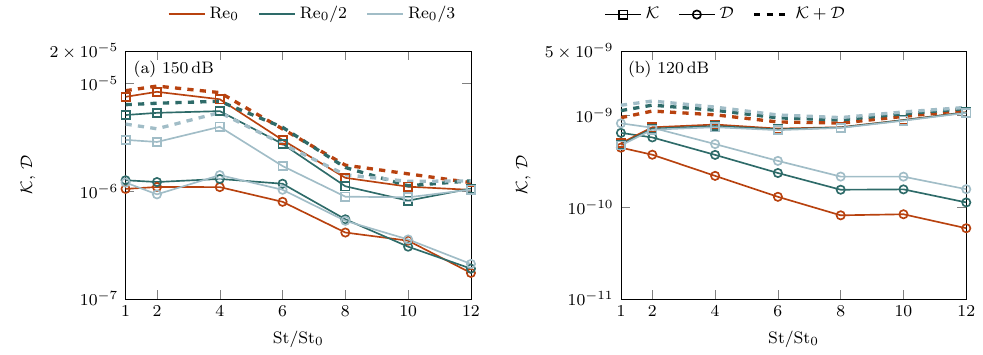}
    \caption{
    Overall kinetic energy, ${\mathcal{K}}$, and viscous loss, ${\mathcal{D}}$, components, obtained by integrating over all frequency components via $\mathcal{S}( \, \cdot \, )$, across $\St$--$\Rey$ combinations for (a) $\mathrm{ISPL} = \SI{150}{\decibel}$ and (b) $\mathrm{ISPL} = \SI{120}{\decibel}$.
    }
    \label{spod_DKE_integral} 
\end{figure}

\Cref{spod_DKE_integral} shows the overall KE and VL obtained by summing the contributions from all frequency components and retaining only the first three leading SPOD modes,
 across the entire parameter space.
At a higher sound pressure level of $\mathrm{ISPL} = \SI{150}{\decibel}$, the total VL accounts for approximately \SIrange[]{20}{60}{\percent} of the total KE.
The combined total energy, ${\mathcal{P}}={\mathcal{K}}+{\mathcal{D}}$, exhibits a three-stage trend that aligns with the absorption coefficient shown in \cref{dns_absorption}.
At low $\St$, the total energy shows little sensitivity to $\Rey$.
For intermediate frequencies, $4\leq\St/\St_0 < 10$, ${\mathcal{P}}$ increases monotonically with $\Rey$.
At higher frequencies, $10 \leq \St/\St_0 \leq 12$, the total spectral energy is nearly independent of $\Rey$, indicating that Reynolds number effects become negligible in this regime because of the absence of significant broadband components.
Among the cases, $\Rey=\Rey_0$ consistently yields the highest total energy conversion across the full range of acoustic frequencies, indicating the strongest energy dissipation mechanism.
At a lower sound pressure level $\mathrm{ISPL} = \SI{120}{\decibel}$, VL plays a more significant role in the overall dissipation mechanism, ranging from approximately \SIrange{20}{180}{\percent} of the KE contribution. 
The $\St$- and $\Rey$-dependence of the spectral KE is weaker than in the higher-amplitude cases, indicating reduced conversion of incident acoustic energy into vortical motion at lower forcing levels.
In contrast, for smaller $\Rey$, VL is larger, highlighting the growing importance of direct viscous attenuation at low ISPL.
The trend of the total energy conversion is consistent with the absorption coefficient shown in \cref{dns_absorption}.

Taken together, these results provide a unified picture for interpreting dissipation mechanisms in acoustic slit flows, specifically those associated with slit resonator mouths.
At low $\St$ and low ISPL, the incident acoustic energy loss receives comparable contributions from the KE of the induced flow field and spatially distributed VL. 
For larger $\St$ and $\mathrm{ISPL}$, the dissipation becomes more localized near the slit mouth, and the KE associated with shed vortices comes to dominate the acoustic damping mechanism.
This evolving balance between viscous and vortex-driven dissipations underlies the nonlinear acoustic damping observed in slit openings at high ISPL and governs energy transfer efficiency and overall damping performance.

\section{Discussion}
\label{sec:Discussion}

We use SPOD to disentangle the energy contributions, KE and VL, in a 2D acoustically driven slit configuration for various frequencies.
We directly link the contributions to their corresponding coherent spatial structures.
This spectral analysis view reveals rich and distinct dissipation behaviors across the parameter space.
The overall VL at $\mathrm{ISPL}=\SI{150}{\decibel}$ remains high, reaches its maximum near $\St = 4\St_0$, and then drops by approximately \SI{50}{\percent} for $\St \ge 6\St_0$.
At this ISPL, dissipation is primarily driven by the conversion of incident acoustic energy into vortical kinetic energy.
We interpret the $\St$-dependence of this behavior through the Keulegan--Carpenter number, $\Kc \equiv U/fd$, where $U$ is the amplitude of the particle velocity oscillation.
For $\mathrm{ISPL}=\SI{150}{\decibel}$ with $\widehat{A} \approx \SI{894}{\pascal}$, we estimate the amplitude using a plane-wave approximation,
\begin{align}\label{eqn: plane_wave_impedance}
    U = \frac{\widehat{A}}{\rho c}.
\end{align}
For $\St = 4\St_0$, this gives $U \approx \SI{2.13}{\meter \per \second}$ and $\Kc \approx 1.33$.
$\Kc$ of order unity implies that the acoustic displacement amplitude is comparable to the slit thickness: fluid particles traverse the entire slit width during each oscillation cycle.
This behavior promotes boundary-layer separation, consistent with the large velocity gradients shown in \cref{spod_BL}, and results in the viscous attenuation of acoustic energy.
Conversely, for $\St > 4\St_0$, the corresponding $\Kc < 1$, so the displacement amplitude is smaller than the slit thickness.

The boundary layers tend to remain attached, and the flow organizes into more confined, X-shaped SPOD modes, as observed in \cref{spod_mode1}, with reduced viscous dissipation.
Previous studies have observed similar results in relevant physical configurations.
Viscous dissipation was found to degrade the performance of sound-driven jet pumps once BL separation sets in, occurring around $\Kc > 0.7$~\citep{oosterhuis2017flow} and $\Kc \approx 1.3$~\citep{timmer2016characterization}.
Likewise, \citet{sarpkaya1986force} reported a maximum drag coefficient for a circular cylinder in oscillatory flow at $\Kc \approx 1$ and $\Rey = 11240$.
At the lower sound pressure level of $\mathrm{ISPL}=\SI{120}{\decibel}$, dissipation is instead dominated by VL, and the overall KE shows a much weaker dependence on $\St$ compared with $\mathrm{ISPL}=\SI{150}{\decibel}$.
For all cases at this ISPL, the small $\Kc$ indicates that particle displacements are much smaller than the slit thickness, and boundary-layer separation is suppressed.
As a result, for larger $\St$, the overall VL is smaller, consistent with weaker shear-driven dissipation in the boundary layers.

We observe that tonal peaks emerge at non-harmonic frequencies with high $\St$, particularly at lower $\Rey$.
The appearance of this spectral branching and the associated non-harmonic peaks indicates that nonlinear triadic interactions play an important role in redistributing energy across frequencies at large $\St$. 
These interactions introduce additional dissipation pathways and are thus important to understand for optimal slit design.
However, such dynamics cannot be directly predicted by linear theory or by linear-based modal decomposition techniques such as POD, DMD, or SPOD, due to the inherently nonlinear nature of the dynamics. 
In the present work, SPOD-based spectral analysis statistically isolates dominant coherent structures from the broadband response but does not resolve the mechanisms governing inter-frequency energy transfer. 
To identify the optimal coherent structures participating in triadic interactions and to quantify the associated energy budgets among these triadic modes, the recently proposed triadic orthogonal decomposition (TOD) by \citet{yeung2024revealing} offers a promising path forward.
By construction, TOD identifies coherent flow structures that optimally represent spectral momentum transfer, quantifies their coupling and energy exchange through an energy-budget bispectrum, and localizes the regions where these interactions occur. 
This analysis is well-suited to regimes where spectral peaks arise at non-integer multiples of the forcing frequency.

Through spectral analysis, the dissipation mechanism we observed offers several implications for the performance of a slit resonator, regardless of whether the application requires high or low damping.
Beyond conventional design based solely on the resonant frequency, energy dissipation can be enhanced by tailoring the slit-mouth thickness to deliberately scale  $\St$, $\Rey$, and $\Kc$, thereby maximizing viscous loss in the desired operating range.
In particular, operating near $\Kc\approx 1$ at high ISPL promotes strong boundary-layer separation and efficient shear-driven attenuation.
At the same time, lower $\Kc$ regimes favor more linear, viscosity-dominated damping with reduced sensitivity to broadband fluctuations.
These findings suggest practical guidelines for acoustic liner and metasurface design in applications such as aircraft engine liners, duct acoustics, and noise-control devices, where robust, broadband, and amplitude-tolerant damping is required.
By explicitly accounting for the coupled roles of $\St$, $\Rey$, ISPL, and $\Kc$, we can engineer slit resonators and related subwavelength elements for optimal impedance at a target frequency.
They can also be designed for controlled nonlinear damping characteristics under realistic high-amplitude operating conditions.

The present study focuses on a simplified two-dimensional slit with anechoic termination, rather than a full slit resonator.
While both systems share the same fundamental mechanism of acoustic--vorticity conversion at the slit mouth, only the resonator can support standing-wave resonance for $f_s \le \SI{6}{\kilo \hertz}$ due to its reflective hard termination.
By removing resonance effects, the simplified slit configuration isolates the local generation and dissipation of vortices at the slit mouth, allowing for the examination of energy-transfer mechanisms without contamination from global resonant modes.
This strategy is canonical: first analyzing elementary configurations, then addressing fully coupled resonant systems.
Accordingly, the present work can be viewed as a first step toward a complete characterization of nonlinear acoustic damping in slit resonators. 
In a companion study, the analysis will be extended to true resonator configurations with hard terminations and discrete resonance frequencies.

\section{Conclusion}
\label{sec:Conclusion}

This study develops rigorous numerical and spectral analyses for quantifying the conversion of incident acoustic energy into coherent, energy-ranked vortical kinetic energy (KE) and viscous loss (VL) in a two-dimensional thin slit.
Direct numerical simulations are performed over a broad parameter space in Strouhal number ($\St$), Reynolds number ($\Rey$), and sound level (ISPL), yielding a dataset with rich and complex dynamics across a range of sound excitations.
Rather than characterizing only the net acoustic energy dissipation via power absorption coefficients, spectral proper orthogonal decomposition (SPOD) is used to separate and quantify, mode-by-mode, the contributions of KE and VL at each frequency.
For each spectral component, the associated KE and VL mechanisms are linked to their coherent spatial structures, revealing robust, physically interpretable processes that govern nonlinear acoustic dissipation within and near slit-type openings.
This analysis reveals that acoustic--KE--VL energy exchange at the slit mouth is governed by the coupled effect of $\St$, $\Rey$, and ISPL. 
Their interplay controls not only the coherence and strength of vortex shedding, but also the partitioning between spectral KE and VL that ultimately sets the dissipation characteristics of the acoustic slit.
The conventional, overall acoustic energy dissipation picture is recovered by summing the KE and VL contributions obtained from the SPOD modes, followed by spatial and spectral summations.

Across all $\St$--$\Rey$--ISPL combinations, more than \SI{99}{\percent} of the VL originates from the near-slit region (within the $16d \times 8d$ region shown in \cref{spod_DKE_2kHz} (b). 
The leading SPOD rank at the fundamental frequency accounts for over \SI{95}{\percent} of the total energy, so the induced flow field is low-rank at the fundamental frequency.
The corresponding dissipation field aligns with regions of maximum velocity gradient, confirming that incident acoustic energy loss is governed primarily by boundary-layer (BL) dynamics and their nonlinear interactions with the slit geometry.

At $\mathrm{ISPL} = \SI{150}{\decibel}$, the largest integrated KE and VL occur near conditions corresponding to a Keulegan--Carpenter number $\Kc \approx 1.33$, where the displacement amplitude is comparable to the slit thickness.
In this regime, the oscillatory BL undergoes separation, generating large-scale vortices that dominate the SPOD spectra and drive strong conversion of incident acoustic energy into KE and VL.
For $\St > 8\St_0$, contraction of the oscillatory BL suppresses separation, yielding more confined, X-shaped modes and an associated $\sim \SI{50}{\percent}$ reduction in dissipation, despite persistent harmonic content.
At lower $\St$ (e.g., $\St = \St_0$), KE-dominated dissipation is less pronounced, and the $\Rey$-dependence of the modal structures becomes more evident.
In this regime, the total VL increases monotonically with decreasing $\Rey$, indicating the growing dominance of direct viscous attenuation.

Case studies at $(\St, \Rey) = (4\St_0, \Rey_0/2)$ and $(8\St_0, \Rey_0)$ show that increasing the slit thickness (i.e., increasing the Womersley number) produces more confined fundamental SPOD modes, sharper harmonic peaks, and reduced broadband KE.
This result demonstrates that thickness-selective design can induce changes in dissipation comparable to those obtained by varying the Strouhal number, $\St$.
Increasing both the slit thickness and $\St$ tends to compress vortex-shedding structures toward the slit mouth, reshaping the balance between tonal and broadband dissipation.

\section*{Acknowledgments}

H.\ Y.\ thanks Profs. Krishan K.\ Ahuja and Lakshmi N.\ Sankar for fruitful discussions. 

\section*{Funding}
H.Y. acknowledges support from the Georgia Institute of Technology Small Bets Internal Research and Development (IRAD) Program.
T.C. and S.H.B. acknowledge support from the U.S.\ Department of Defense, the Army Research Office under Grant No. W911NF-23-10324 (PMs Drs.\ Denise Ford and Robert Martin). 
This work used NCSA Delta through allocation TG-PHY210084 (PI~Bryngelson) from the Advanced Cyberinfrastructure Coordination Ecosystem: Services \& Support (ACCESS) program, supported by National Science Foundation grants \#2138259, \#2138286, \#2138307, \#2137603, and \#2138296.
We acknowledge the use of computational resources at Georgia~Tech, including PACE~Phoenix.

\section*{Declaration of competing interest}

The authors declare that they have no known competing financial interests or personal relationships that could have appeared to influence the work reported in this paper.

\section*{Statement of data availability}

MFC is available in perpetuity under the MIT license at \url{github.com/MFlowCode/MFC}.
The database is permanently available in the Georgia Tech Digital Repository at \href{https://www.dropbox.com/scl/fo/kbyq6lu2jm15ouy6qyaxo/ALW6qPc5-x_jZvy_6bFOQA0?rlkey=t5quat3iqzxuq8j78je2706gi&st=5csji168&dl=0}{\tt this link} and this reference~\citep{database}.

\section*{Author ORCIDs}

H.~Yu, \url{https://orcid.org/0009-0002-1388-5330};\\
T.~Chu, \url{https://orcid.org/0000-0001-8587-7925};\\
S.~H.~Bryngelson, \url{https://orcid.org/0000-0003-1750-7265}.

\section*{CRediT authorship contribution statement}

\textbf{HY}: Conceptualization, Formal analysis, Methodology, Software, Investigation, Data Curation, Validation, Visualization, Writing -- original draft, Writing -- review $\&$ editing.
\textbf{TC}: Conceptualization, Formal analysis, Methodology, Software, Investigation, Validation, Visualization, Supervision, Writing -- original draft, Writing -- review $\&$ editing.
\textbf{SHB}: Conceptualization, Funding acquisition, Methodology, Project administration, Resources, Supervision, Writing -- original draft, Writing -- review $\&$ editing.

\appendix
\setcounter{equation}{0}
 \renewcommand{\theequation}{{\rm A}.\arabic{equation}}
 \setcounter{table}{0}
\renewcommand{\thetable}{{\rm A}.\arabic{table}}
\setcounter{figure}{0}
 \renewcommand{\thefigure}{{\rm A}.\arabic{figure}}

\section*{Appendix A. Additional high ISPL spectral results}
\label{appendixA}

In this appendix, we include the fundamental modes, SPOD eigenvalues, and dissipation spectra of the studied cases with $\St=2\St_0$, $=6\St_0$, and $=10 \St_0$ at $\mathrm{ISPL} = \SI{150}{\decibel}$. 
These cases were omitted from the main text in \cref{sec:spectral} for brevity but are included here for completeness, as they further illustrate the evolution of the modal structure, spectral energy distribution, and dissipation across the full range of parameters investigated.
\begin{figure}[!ht]
    \centering
    \includegraphics[]{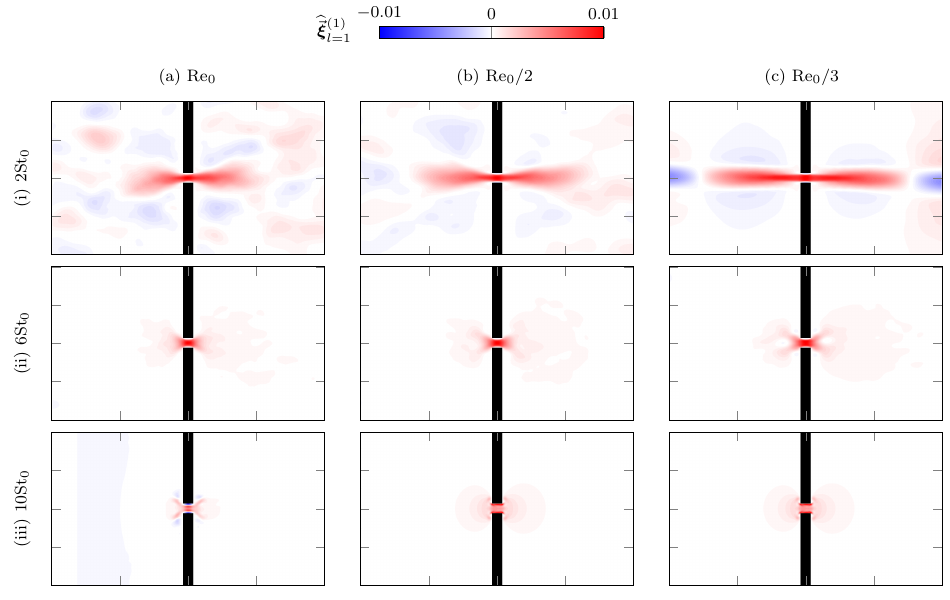}
    \caption{
     Leading SPOD modes, $\smash{\widehat{\vec{\bm{\mathrm{\xi}}}}}_{l=1}^{(1)}$, for the complementary $\St$--$\Rey$ combinations at $\mathrm{ISPL} = \SI{150}{\decibel}$ shown in \cref{spod_mode1}, plotted at their respective fundamental frequencies ($l=1$) as their $x$-directional components, $u$.
    Cases in column (a)--(c) represent $\Rey=\Rey_0$, $\Rey_0/2$, and $\Rey_0/3$. Cases in row (i)--(iii) represent $\St=2\St_0$, $6\St_0$,and $10\St_0$.
    }
    \label{spod_mode_restcases} 
\end{figure}

\Cref{spod_mode_restcases} displays the fundamental modes of these cases.
At $\St = 2\St_0 $, the fundamental mode retains relatively large-scale structures 
Similar to cases with $\St \le 4\St_0$, the coherent structure extends outward from the slit further with $\Rey$ reduced.
At $\St \ge 6\St_0$,  the fundamental modes exhibit X-shaped patterns near the slit, similar to cases at $\St = 8\St_0$ and $=12\St_0$.

\begin{figure}[ht!]
    \centering
    \includegraphics[]{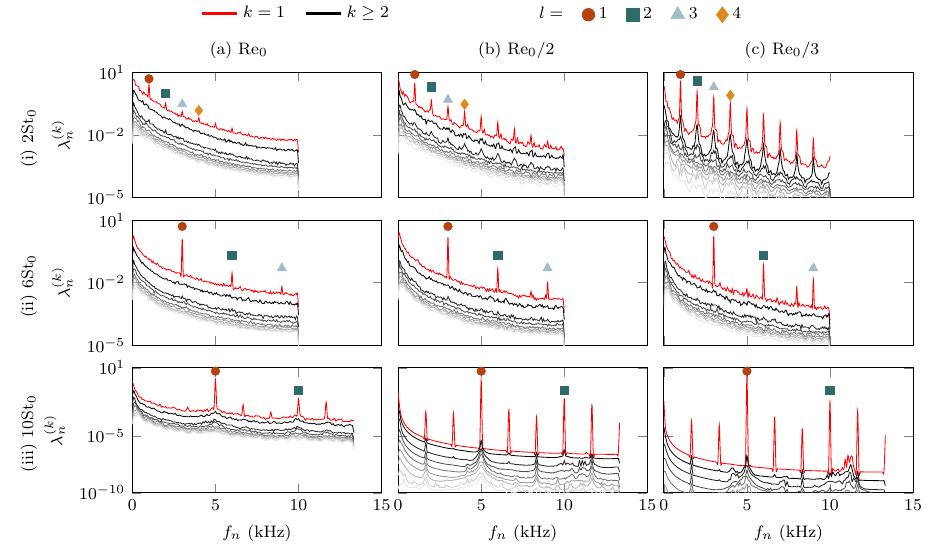}
    \caption{
    SPOD eigenvalue spectra, $\lambda_n^{(k)}$, at different rank $k$ for the complementary $\St$--$\Rey$ combinations at  $\mathrm{ISPL} = \SI{150}{\decibel}$ shown in \cref{spod_eigenvalue}.
    Cases in column (a)--(c) represent $\Rey=\Rey_0$, $\Rey_0/2$, and $\Rey_0/3$. Cases in row (i)--(iii) represent $\St=2\St_0$, $6\St_0$,and $10\St_0$.
    The fundamental frequency ($l=1$ defined in \cref{eqn:harmonics}) and its higher-order harmonics ($l \geq 2$) manifest as distinct peaks in the spectrum.
    For example, two tonal peaks arise at $f_n = \SI{5}{\kilo \hertz}$ ($l=1$) and $f_n = \SI{10}{\kilo \hertz}$ ($l=2$) in case (a, iii).
    }
    \label{spod_eigenvalue_restcases} 
\end{figure}

\begin{figure}[ht!]
    \centering
    \includegraphics[]{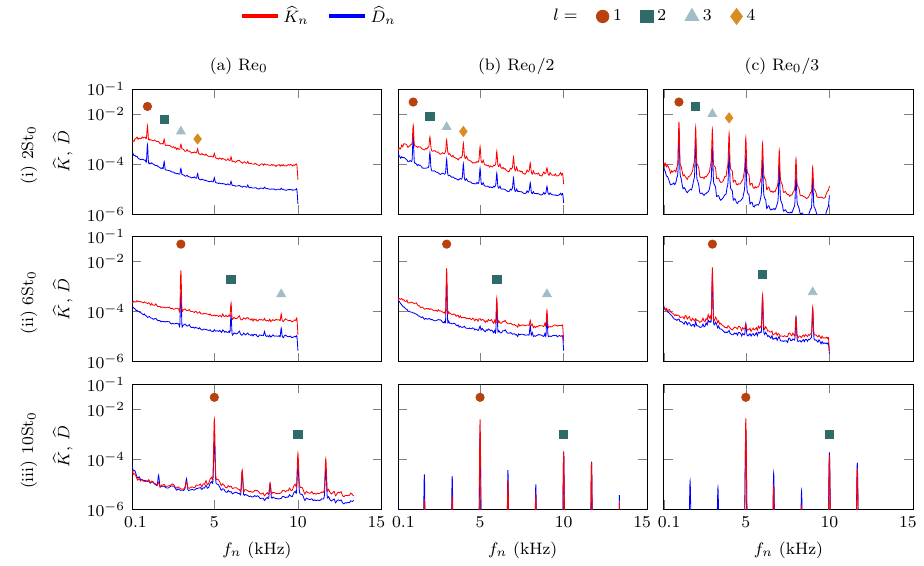}
    \caption{        
    Viscous loss (VL), $\widehat{D}_n$, and frequency-weighted kinetic energy (KE) spectra, $\widehat{K}_n$, for the complementary $\St$--$\Rey$ combinations at $\mathrm{ISPL} = \SI{150}{\decibel}$ shown in \cref{spod_DKE_spectra}.
    Cases in column (a)--(c) represent $\Rey=\Rey_0$, $\Rey_0/2$, and $\Rey_0/3$. Cases in row (i)--(iii) represent $\St=2\St_0$, $6\St_0$,and $10\St_0$.
    }
    \label{spod_DKE_restcases} 
\end{figure}

\Cref{spod_eigenvalue_restcases} displays the SPOD eigenvalues of these cases.
As $\St$ increases and $\Rey$ decreases, the spectral peaks sharpen and the rank hierarchy becomes more distinct.
The reduction of broadband energy with decreasing $\Rey$ is also evident across all three $\St$, consistent with the results observed in \cref{subsec:parameter_spectrum} with lower $\Rey$ flows.

\Cref{spod_DKE_restcases} displays the dissipation spectra of these cases.
Across all three $\St$, the spectral VL peaks are aligned with the peaks of the KE spectra, indicating that viscous losses remain closely tied to the vorticity dynamics of the dominant coherent mode.
With larger $\St$ and smaller $\Rey$, the difference between the total KE and VL diminishes noticeably, consistent with reduced vorticity coherence and increased viscosity.

\section*{Appendix B. Low ISPL spectral results}
\label{appendixB}

\begin{figure}[!ht]
    \centering
    \includegraphics[]{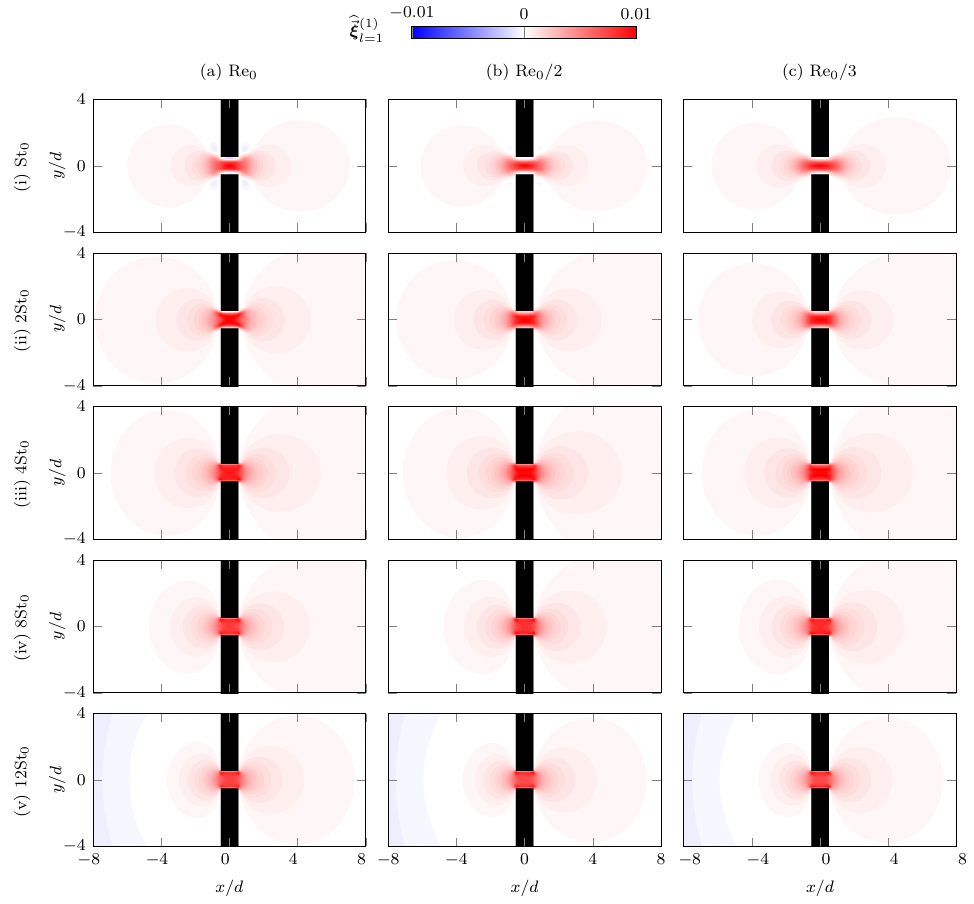}
    \caption{
     Leading SPOD modes, $\smash{\widehat{\vec{\bm{\mathrm{\xi}}}}}_{l=1}^{(1)}$, for selected $\St$--$\Rey$ combinations at $\mathrm{ISPL} = \SI{120}{\decibel}$, plotted at their respective fundamental frequencies ($l=1$) as their $x$-directional components, $u$.
    Cases in column (a)--(c) represent $\Rey=\Rey_0$, $\Rey_0/2$, and $\Rey_0/3$. Cases in row (i)--(v) represent $\St=\St_0$, $2\St_0$, $4\St_0$, $8\St_0$, ,and $12\St_0$.
    }
    \label{spod_mode1_120dB} 
\end{figure}

In this appendix, we include the modes at the fundamental frequency and the dissipation spectra of the selected studied cases at $\mathrm{ISPL} = \SI{120}{\decibel}$, which are not present in \cref{sec:spectral} for brevity.
\Cref{spod_mode1_120dB} displays the fundamental modes of these cases.
For $\St = \St_0$, we observed dumbbell-shaped structures with thicker and separated BL near the slit wall.
The separation is apparent at $\Rey = \Rey_0$ and is weaker for $\Rey \le \Rey_0/2$.
For $\St = 2\St_0$, the structure appears to be an intermediate shape between dumbbell-shaped and X-shaped, with barely distinguishable $\Rey$-dependence.
When $\St \ge 4\St_0$, the modal structures look similar across different $\Rey$, indicating a negligible difference in the coherent vortical structure across the parameter space.

\begin{figure}[ht!]
    \centering
    \includegraphics[]{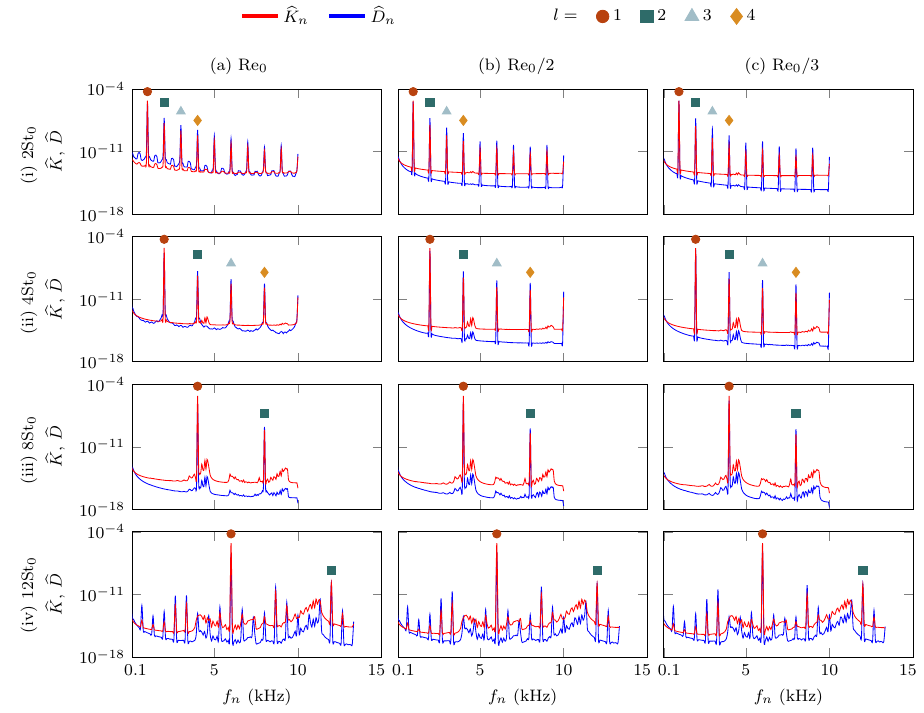}
    \caption{
        Viscous loss (VL), $\widehat{D}_n$, and frequency-weighted kinetic energy (KE) spectra, $\widehat{K}_n$,
        for the complementary $\St$--$\Rey$ combinations at $\mathrm{ISPL} = \SI{120}{\decibel}$ shown in \cref{spod_DKE_120dB}.
        The fundamental frequency ($l=1$ defined in \cref{eqn:harmonics}) and its higher-order harmonics ($l \geq 2$) are manifested as distinct peaks in the spectrum.
        Cases in column (a)--(c) represent $\Rey=\Rey_0$, $\Rey_0/2$, and $\Rey_0/3$. Cases in row (i)--(iV) represent $\St=2\St_0$, $4\St_0$, $8\St_0$, and $12\St_0$.
    }
    \label{spod_DKE_spectra_120dB_full} 
\end{figure}

\Cref{spod_DKE_spectra_120dB_full} displays the dissipation spectra of these cases.
The broadband spectral energy is at least five orders of magnitude smaller than the narrow tonal peaks across all the parameter space.
So, the broadband contribution of the KE is negligible at this ISPL.
Additional peaks at non-integer multiples of the forcing frequency appear at $\St \ge 4\St_0$, suggesting nonlinear interactions with the slit boundaries even at an $\mathrm{ISPL}$ as low as $\SI{120}{\decibel}$.

\section*{Appendix C. Compressibility contribution of the dissipation}
\label{appendixC}

\begin{figure}[ht!]
    \centering
    \includegraphics[]{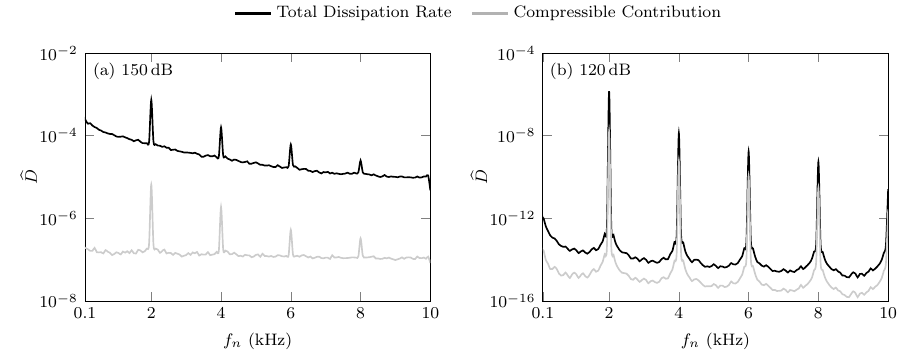}
    \caption{
        Total and compressible viscous loss (VL), $ \widehat{D}_n$, contribution spectra at $\St=4\St_0$ and $\Rey=\Rey_0$, subject to sound level at (a)~$\mathrm{ISPL} = \SI{150}{\decibel}$ and (b)~$\mathrm{ISPL} = \SI{120}{\decibel}$.
        The compressible contributions are less than \SI{1}{\percent} of the total viscous dissipation to justify the incompressible assumption in the spectral analysis.
    }
    \label{spod_compressibility} 
\end{figure}

\Cref{spod_compressibility} shows spectra of both contributions at the representative case demonstrated in \cref{subsec:representative_spod}.
The deviatoric contribution is two orders of magnitude larger than the compressible contribution at both $\mathrm{ISPL} = \SI{150}{\decibel}$ and $\mathrm{ISPL} = \SI{120}{\decibel}$ conditions.
This result justifies the incompressible assumption made in \cref{subsec:Database_setup}.

\clearpage

\bibliography{main.bib}

@article{Gottlieb1998,
    title        = {Total variation diminishing {R}unge--{K}utta schemes},
    author       = {Gottlieb, Sigal and Shu, Chi-Wang},
    year         = 1998,
    journal      = {Math. Comput.},
    volume       = 67,
    number       = 221,
    pages        = {73--85}
}

@article{thompson1990,
    title        = {Time-dependent boundary conditions for hyperbolic systems, {II}},
    author       = {Kevin W Thompson},
    year         = 1990,
    journal      = {J. Comput. Phys.},
    volume       = 89,
    number       = 2,
    pages        = {439--461},
}

@article{tseng2003ghost,
    title        = {A ghost-cell immersed boundary method for flow in complex geometry},
    author       = {Tseng, Yu-Heng and Ferziger, Joel H},
    year         = 2003,
    journal      = {J. Comput. Phys.},
    volume       = 192,
    number       = 2,
    pages        = {593--623},
}

@article{cummings1984acoustic,
  title={Acoustic nonlinearities and power losses at orifices},
  author={Cummings, A},
  year={1984},
  journal={AIAA J.},
  volume={22},
  number={6},
  pages={786--792},
}

@article{howe1980dissipation,
  title={The dissipation of sound at an edge},
  author={Howe, Michael S},
  year={1980},
  journal={J. Sound Vib.},
  volume={70},
  number={3},
  pages={407--411},
}

@inproceedings{yu2024numerical,
  title        ={Numerical investigation of leakage of high-Amplitude sound in ill-Fitting earplugs},
  author       ={Yu, H. and Ahuja, K. K. and Sankar, L. N. and Bryngelson, S. H.},
  booktitle    ={AIAA AVIATION},
  year         = 2024,
  pages={4391},
}

@article{yu2025transmission,
  title        ={Transmission of high-Amplitude sound through leakages of ill-fitting earplugs},
  author       ={Yu, H. and Ahuja, K. K. and Sankar, L. N. and Bryngelson, S. H.},
  year         = 2025,
  journal      ={arXiv preprint arXiv:2510.16355},
}

@article{wilfong2025mfc,
  title        = {{MFC 5.0: A}n exascale many-physics flow solver},
  author       = {Wilfong, Benjamin and Berre, Henry A Le and Radhakrishnan, Anand and Gupta, Ansh and Vaca-Revelo, Diego and Adam, Dimitrios and Yu, Haocheng and Lee, Hyeoksu and Chreim, Jose Rodolfo and Barbosa, Mirelys Carcana and others},
  year         = 2025,
  journal      = {arXiv preprint arXiv:2503.07953},
}

@article{bryngelson2021mfc,
  title        = {{MFC: A}n open-source high-order multi-component, multi-phase, and multi-scale compressible flow solver},
  author       = {Bryngelson, S. H. and Schmidmayer, K. and Coralic, V. and Maeda, K. and Meng, J. and Colonius, T.},
  year         = 2021,
  journal      = {Comput. Phys. Commun.},
  volume       = 266,
  pages        = 107396,
}

@article{tang2023combined,
  title        = {A combined delayed detached eddy simulation and linearized {N}avier--{S}tokes equation study on the generation and reduction of aerodynamic noises inside steam turbine control valve with acoustic liner},
  author       = {Tang, Yuchao and Wang, Peng and Liu, Yingzheng},
  year         = 2023,
  journal      = {J. Fluids Eng.},
  volume       = 145,
  number       = 12,
  pages        = 121202,
}

@article{tam2010computational,
  title        = {A computational and experimental study of resonators in three dimensions},
  author       = {Tam, Christopher KW and Ju, Hongbin and Jones, Michael G and Watson, Willie R and Parrott, Tony L},
  year         = 2010,
  journal      = {J. Sound Vib.},
  volume       = 329,
  number       = 24,
  pages        = {5164--5193},
}

@article{tam2005computational,
  title        = {A computational and experimental study of slit resonators},
  author       = {Tam, Christopher KW and Ju, Hongbin and Jones, Michael G and Watson, Willie R and Parrott, Tony L},
  year         = 2005,
  journal      = {J. Sound Vib.},
  volume       = 284,
  number       = {3-5},
  pages        = {947--984},
}

@article{tam2001numerical,
  title        = {A numerical and experimental investigation of the dissipation mechanisms of resonant acoustic liners},
  author       = {Tam, Christopher KW and Kurbatskii, Konstantin A and Ahuja, KK and Gaeta Jr, RJ},
  year         = 2001,
  journal      = {J. Sound Vib.},
  volume       = 245,
  number       = 3,
  pages        = {545--557},
}

@article{maeda2017source,
  title        = {A source term approach for generation of one-way acoustic waves in the Euler and {N}avier--{S}tokes equations},
  author       = {Maeda, Kazuki and Colonius, Tim},
  year         = 2017,
  journal      = {Wave Motion},
  volume       = 75,
  pages        = {36--49},
}

@article{bugeat2024acoustic,
  title        = {Acoustic resolvent analysis of turbulent jets},
  author       = {Bugeat, B. and Karban, U. and Agarwal, A. and Lesshafft, L. and Jordan, P.},
  year         = 2024,
  journal      = {Theor. Comput. Fluid Dyn.},
  volume       = 38,
  number       = 5,
  pages        = {687--706},
}

@article{qiang2022aeroacoustic,
  title        = {Aeroacoustic simulation of transient vortex dynamics subjected to high-intensity acoustic waves},
  author       = {Qiang, Xu and Wang, Peng and Liu, Yingzheng},
  year         = 2022,
  journal      = {Phys. Fluids},
  volume       = 34,
  number       = 9,
}

@article{wang2013cascade,
  title        = {Cascade of kinetic energy in three-dimensional compressible turbulence},
  author       = {Wang, Jianchun and Yang, Yantao and Shi, Yipeng and Xiao, Zuoli and He, XT and Chen, Shiyi},
  year         = 2013,
  journal      = {Phys. Rev. Lett.},
  volume       = 110,
  number       = 21,
  pages        = 214505,
}

@article{timmer2016characterization,
  title        = {Characterization and reduction of flow separation in jet pumps for laminar oscillatory flows},
  author       = {Timmer, Michael AG and Oosterhuis, Joris P and B{\"u}hler, Simon and Wilcox, Douglas and van der Meer, Theo H},
  year         = 2016,
  journal      = {J. Acoust. Soc. Am.},
  volume       = 139,
  number       = 1,
  pages        = {193--203},
}

@article{awasthi2025coherent,
  title        = {Coherent oscillations and acoustic waves in a supersonic cylinder wake},
  author       = {Awasthi, M and McCreton, S and Moreau, DJ and Doolan, CJ},
  year         = 2025,
  journal      = {J. Fluid Mech.},
  volume       = 1012,
  pages        = {A27},
}

@article{schmid2010dynamic,
  title        = {Dynamic mode decomposition of numerical and experimental data},
  author       = {Schmid, Peter J},
  year         = 2010,
  journal      = {J. Fluid Mech.},
  volume       = 656,
  pages        = {5--28},
}

@article{kida1990energy,
  title        = {Energy and spectral dynamics in forced compressible turbulence},
  author       = {Kida, S. and Orszag, S. A.},
  year         = 1990,
  journal      = {J. Sci. Comput.},
  volume       = 5,
  pages        = {85--125},
}

@article{coralic2014finite,
  title        = {Finite-volume {WENO} scheme for viscous compressible multicomponent flows},
  author       = {Coralic, Vedran and Colonius, Tim},
  year         = 2014,
  journal      = {J. Comput. Phys.},
  volume       = 274,
  pages        = {95--121},
}

@article{oosterhuis2017flow,
  title        = {Flow separation and turbulence in jet pumps for thermoacoustic applications},
  author       = {Oosterhuis, Joris P and Verbeek, Anton A and B{\"u}hler, Simon and Wilcox, Douglas and van der Meer, Theo H},
  year         = 2017,
  journal      = {Flow Turbul. Combust.},
  volume       = 98,
  number       = 1,
  pages        = {311--326},
}

@book{landau1987fluid,
  title        = {Fluid {M}echanics},
  author       = {Landau, Lev Davidovich and Lifshitz, Evgeny Mikhailovich},
  year         = 1987,
  publisher    = {Elsevier},
  volume       = 6,
}

@article{sarpkaya1986force,
  title        = {Force on a circular cylinder in viscous oscillatory flow at low {K}eulegan--{C}arpenter numbers},
  author       = {Sarpkaya, Turgut},
  year         = 1986,
  journal      = {J. Fluid Mech.},
  volume       = 165,
  pages        = {61--71},
}

@article{nekkanti2021frequency,
  title        = {Frequency--time analysis, low-rank reconstruction and denoising of turbulent flows using {SPOD}},
  author       = {Nekkanti, Akhil and Schmidt, Oliver T},
  year         = 2021,
  journal      = {J. Fluid Mech.},
  volume       = 926,
  pages        = {A26},
}

@article{schmidt2020guide,
  title        = {Guide to spectral proper orthogonal decomposition},
  author       = {Schmidt, Oliver T and Colonius, Tim},
  year         = 2020,
  journal      = {AIAA J.},
  volume       = 58,
  number       = 3,
  pages        = {1023--1033},
}

@article{leung2007duct,
  title        = {In-duct orifice and its effect on sound absorption},
  author       = {Leung, RCK and So, RMC and Wang, MH and Li, XM},
  year         = 2007,
  journal      = {J. Sound Vib.},
  volume       = 299,
  number       = {4-5},
  pages        = {990--1004},
}

@article{radhakrishnan2024method,
  title        = {Method for scalable and performant {GPU}-accelerated simulation of multiphase compressible flow},
  author       = {Radhakrishnan, Anand and Le Berre, Henry and Wilfong, Benjamin and Spratt, Jean-Sebastien and Rodriguez Jr, Mauro and Colonius, Tim and Bryngelson, Spencer H},
  year         = 2024,
  journal      = {Comput. Phys. Commun.},
  volume       = 302,
  pages        = 109238,
}

@article{womersley1955method,
  title        = {Method for the calculation of velocity, rate of flow and viscous drag in arteries when the pressure gradient is known},
  author       = {Womersley, John R},
  year         = 1955,
  journal      = {J. Physiol.},
  volume       = 127,
  number       = 3,
  pages        = 553,
}

@article{tam2000microfluid,
  title        = {Microfluid dynamics and acoustics of resonant liners},
  author       = {Tam, Christopher KW and Kurbatskii, Konstantin A},
  year         = 2000,
  journal      = {AIAA J.},
  volume       = 38,
  number       = 8,
  pages        = {1331--1339},
}

@article{nekkanti2021modal,
  title        = {Modal analysis of acoustic directivity in turbulent jets},
  author       = {Nekkanti, A. and Schmidt, O. T.},
  year         = 2021,
  journal      = {AIAA J.},
  volume       = 59,
  number       = 1,
  pages        = {228--239},
}

@article{tam2008numerical,
  title        = {Numerical simulation of a slit resonator in a grazing flow under acoustic excitation},
  author       = {Tam, Christopher KW and Ju, Hongbin and Walker, Bruce E},
  year         = 2008,
  journal      = {J. Sound Vib.},
  volume       = 313,
  number       = {3-5},
  pages        = {449--471},
}

@article{ingard1953theory,
  title        = {On the theory and design of acoustic resonators},
  author       = {Ingard, Uno},
  year         = 1953,
  journal      = {J. Acoust. Soc. Am.},
  volume       = 25,
  number       = 6,
  pages        = {1037--1061},
}

@article{tang2025phase,
  title        = {Phase-locking {PIV} measurement of vortex--vortex interactions inside dual-slit cavity during high-intensity acoustic modulation},
  author       = {Tang, Yuchao and Wang, Peng and Liu, Yingzheng},
  year         = 2025,
  journal      = {Exp. Therm. Fluid Sci.},
  volume       = 166,
  pages        = 111483,
}

@article{yeung2024revealing,
  title        = {Revealing structure and symmetry of nonlinearity in natural and engineering flows},
  author       = {Yeung, Brandon and Chu, Tianyi and Schmidt, Oliver T},
  year         = 2024,
  journal      = {arXiv preprint arXiv:2411.12057},
}

@article{schmidt2018spectral,
  title        = {Spectral analysis of jet turbulence},
  author       = {Schmidt, O. T. and Towne, A. and Rigas, G. and Colonius, T. and Br{\`e}s, G. A.},
  year         = 2018,
  journal      = {J. Fluid Mech.},
  volume       = 855,
  pages        = {953--982},
}

@article{towne2018spectral,
  title        = {Spectral proper orthogonal decomposition and its relationship to dynamic mode decomposition and resolvent analysis},
  author       = {Towne, Aaron and Schmidt, Oliver T and Colonius, Tim},
  year         = 2018,
  journal      = {J. Fluid Mech.},
  volume       = 847,
  pages        = {821--867},
}

@article{himeno2021spod,
  title        = {{SPOD} analysis of noise-generating {R}ossiter modes in a slat with and without a bulb seal},
  author       = {Himeno, Fernando HT and Souza, Daniel S and Amaral, Filipe R and Rodr{\'\i}guez, Daniel and Medeiros, Marcello AF},
  year         = 2021,
  journal      = {J. Fluid Mech.},
  volume       = 915,
  pages        = {A67},
}

@article{chu2025stochastic,
  title        = {Stochastic reduced-order {K}oopman model for turbulent flows},
  author       = {Chu, Tianyi and Schmidt, Oliver T},
  year         = 2025,
  journal      = {Proc. R. Soc. Lond. A},
  volume       = 481,
  number       = 2323,
  pages        = 20250270,
}

@book{lumley1970stochastic,
  title        = {Stochastic {T}ools in {T}urbulence},
  author       = {Lumley, J. L.},
  year         = 1970,
  publisher    = {Academic Press},
  address      = {New York},
}

@article{berkooz1993proper,
  title        = {The proper orthogonal decomposition in the analysis of turbulent flows},
  author       = {Berkooz, Gal and Holmes, Philip and Lumley, John L},
  year         = 1993,
  journal      = {Annu. Rev. Fluid Mech.},
  volume       = 25,
  number       = 1,
  pages        = {539--575},
}

@article{lumley1967structure,
  title        = {The structure of inhomogeneous turbulent flows},
  author       = {Lumley, J. L.},
  year         = 1967,
  journal      = {Atm. Turbul. Radio Wave Propag.},
}

@article{welch1967use,
  title        = {The use of fast {F}ourier transform for the estimation of power spectra: {A} method based on time averaging over short, modified periodograms},
  author       = {Welch, P.},
  year         = 1967,
  journal      = {IEEE Trans. Audio Electroacoust.},
  volume       = 15,
  number       = 2,
  pages        = {70--73},
}

@article{sano2019trailing,
  title        = {Trailing-edge noise from the scattering of spanwise-coherent structures},
  author       = {Sano, Alex and Abreu, Leandra I and Cavalieri, Andr{\'e} VG and Wolf, William R},
  year         = 2019,
  journal      = {Phys. Rev. Fluids},
  volume       = 4,
  number       = 9,
  pages        = {094602},
}

@article{chung1980transfer,
  title        = {Transfer function method of measuring in-duct acoustic properties. {I}. {T}heory},
  author       = {Chung, JY and Blaser, DA},
  year         = 1980,
  journal      = {J. Acoust. Soc. Am.},
  volume       = 68,
  number       = 3,
  pages        = {907--913},
}

@article{sirovich1987turbulence,
  title        = {Turbulence and the dynamics of coherent structures. {I}. {C}oherent structures},
  author       = {Sirovich, L.},
  year         = 1987,
  journal      = {Q. Appl. Math.},
  volume       = 45,
  number       = 3,
  pages        = {561--571},
}

@book{white1991viscous,
  title        = {Viscous {F}luid {F}low 2nd {E}dition},
  author       = {White, FM},
  year         = 1991,
  publisher    = {{M}c{G}raw-{H}ill},
  address = {New York},
}

@article{nogueira2022wave,
  title        = {Wave-packet modulation in shock-containing jets},
  author       = {Nogueira, P. A. S. and Self, H. W. A. and Towne, A. and Edgington-Mitchell, D.},
  year         = 2022,
  journal      = {Phys. Rev. Fluids},
  volume       = 7,
  number       = 7,
  pages        = {074608},
}

@article{abily2023non,
  title={Non-linear effects in thin slits for low frequency sound absorption},
  author={Abily, Thibault and Regnard, Josselin and Gabard, Gw{\'e}na{\"e}l and Durand, St{\'e}phane},
  journal={J. Sound Vib.},
  volume={546},
  pages={117432},
  year={2023},
}

@article{chen2020microscopic,
  title={Microscopic fluid dynamics of a wire screen bound to a slit resonator excited by acoustic waves},
  author={Chen, C and Li, XD},
  journal={Phys. Fluids},
  volume={32},
  number={11},
  year={2020},
}

@article{aulitto2022effect,
  title={Effect of slit length on linear and non-linear acoustic transfer impedance of a micro-slit plate},
  author={Aulitto, Alessia and Hirschberg, Avraham and Arteaga, Ines Lopez and Buijssen, Esm{\'e}e LRH},
  journal={Acta. Acust},
  volume={6},
  pages={6},
  year={2022},
}

@article{qu2023broadband,
  title={Broadband quasi-perfect sound absorption by a metasurface with coupled resonators at both low-and high-amplitude excitations},
  author={Qu, Renhao and Guo, Jingwen and Fang, Yi and Zhong, Siyang and Zhang, Xin},
  journal={Mech. Syst. Signal Process.},
  volume={204},
  pages={110782},
  year={2023},
}

@article{hoppen2023helmholtz,
  title={Helmholtz resonator with two resonance frequencies by coupling with a mechanical resonator},
  author={Hoppen, Hannah and Langfeldt, Felix and Gleine, Wolfgang and Von Estorff, Otto},
  journal={J. Sound Vib.},
  volume={559},
  pages={117747},
  year={2023},
}

@inproceedings{wilfongGB25,
    author = {Wilfong, Benjamin and Radhakrishnan, Anand and Le Berre, Henry and Vickers, Daniel and Prathi, Tanush and Tselepidis, Nikolaos and Dorschner, Benedikt and Budiardja, Reuben and Cornille, Brian and Abbott, Stephen and Sch\"{a}fer, Florian and Bryngelson, Spencer},
    title = {Simulating many-engine spacecraft: {E}xceeding 1 quadrillion degrees of freedom via information geometric regularization},
    year = {2025},
    booktitle = {SC25},
    pages = {14-24},
}

@inproceedings{wilfong-hpctests,
  title = {Testing and benchmarking emerging supercomputers via the {MFC} flow solver},
  author = {Wilfong, B. and Radhakrishnan, A. and {Le Berre}, H. A. and Prathi, T. and Abbott, S. and Bryngelson, S. H.},
  booktitle = {SC25 Workshops},
  year = {2025},
}

@misc{database,
  author = {H. Yu and T. Chu and S. H. Bryngelson},
  title = {Acoustically-driven slit {DNS} database},
  howpublished = {\url{https://www.dropbox.com/scl/fo/kbyq6lu2jm15ouy6qyaxo/ALW6qPc5-x_jZvy_6bFOQA0?rlkey=t5quat3iqzxuq8j78je2706gi&st=5csji168&dl=0}},
  year = {2025},
  note = {Accessed: 2025-12-13}
}

\end{document}